\documentclass[%
 reprint,
 amsmath,amssymb,
 aps,
]{revtex4-2}

\usepackage{graphicx}
\usepackage{dcolumn}
\usepackage{bm}
\usepackage{color}
\newcommand{\SU}[1]{\ensuremath{\mathrm{SU}( #1 )}}
\newcommand{\SpR}[1]{\ensuremath{\mathrm{Sp}( #1,\mathbb{R} )}}

\newcommand{\hw}{$\hbar\Omega$}

\begin{document}

\preprint{APS/123-QED}

\title{Uncertainty Quantification of Collective Nuclear Observables\\ From the Chiral Potential Parametrization}

\author{Kevin S. Becker}
 \email{kbeck13@lsu.edu}
 \affiliation{Department of Physics and Astronomy, Louisiana State University, Baton Rouge, LA 70803, USA}%
\author{Kristina D. Launey}%
 \affiliation{Department of Physics and Astronomy, Louisiana State University, Baton Rouge, LA 70803, USA}%
\author{Andreas Ekström}
 \affiliation{Department of Physics, Chalmers Institute of Technology, Gothenburg, Sweden}%
\author{Tomáš Dytrych}
 \affiliation{Nuclear Physics Institute of the Czech Academy of Sciences, 250 68 \v{R}e\v{z}, Czech Republic}%
\author{Daniel Langr}
 \affiliation{Department of Computer Systems, Faculty of Information Technology, Czech Technical University in Prague, Prague 16000, Czech Republic}%
\author{Grigor H. Sargsyan}
\affiliation{Facility for Rare Isotope Beams, Michigan State University, East Lansing, MI 48824, USA}
\author{Jerry P. Draayer}%
 \affiliation{Department of Physics and Astronomy, Louisiana State University, Baton Rouge, LA\, 70803, USA}%

\date{\today}

\begin{abstract}
We perform an uncertainty estimate of quadrupole moments and $B(E2)$ transition rates that inform nuclear collectivity. In particular, we study the low-lying states of $^6$Li and $^{12}$C using the \textit{ab initio} symmetry-adapted no-core shell model. For a narrow standard deviation of approximately $1\%$ on the low-energy constants which parametrize high-precision chiral potentials, we find output standard deviations in the collective observables ranging from approximately $3$-$6\%$. The results mark the first step towards a rigorous uncertainty quantification of collectivity in nuclei that aims to account for all sources of uncertainty in \textit{ab initio} descriptions of challenging collective and clustering observables.
\end{abstract}

\maketitle


\section{\label{intro}Introduction}
In advancing the frontier of \textit{ab initio} nuclear modelling and informing the next generation of rare isotope beam experiments, it is critical to quantify uncertainties in the underlying theoretical framework and fully propagate them to observable predictions (for recent efforts, see, e.g., \cite{Wesolowski_2019,odell2023rose,PhysRevLett.122.232502,nat_bmm}). It is often difficult to identify all possible sources of model uncertainty, which for first-principle descriptions can be divided in two general categories. There are those arising from the controlled approximations of the employed quantum many-body framework, such as the error in truncating the (infinitely large) solution space \cite{Wendt_2015,K_nig_2014,PhysRevC.86.054002}. There is also uncertainty in the nuclear Hamiltonian used within this scheme that emerges from the description of the underlying nuclear interaction, typically a chiral potential derived in the effective-field-theory (EFT) framework, and in particular, from the probability distributions of its parameters fitted to data and from the error in truncating the infinite effective series \cite{Furnstahl_2015,PhysRevC.104.064001,Svensson_2023,svensson2023inference}. Even when all sources are properly accounted for, performing the uncertainty analysis is computationally intensive, since it involves sampling high-dimensional distributions of model inputs and performing computationally intensive matrix diagonalizations. Robust uncertainty quantification therefore poses a series of nontrivial theoretical and computational challenges which call for efficient statistical tools and physically informed techniques to lower the computational load.

In this study, we provide an uncertainty estimate on collective observables of light nuclei computed with realistic chiral nucleon-nucleon (NN) interactions (for a review, see, e.g., \cite{Machleidt_2011}) in the symmetry-adapted no-core shell model (SA-NCSM) \cite{DytrychSBDV_PRL07,LauneyDD16,DytrychLDRWRBB20}. Specifically, we study how the $Q_2$ electric quadrupole moments and $B(E2)$ transition rates of low-lying states of $^6$Li and $^{12}$C are impacted by uncertainties in the values of the so-called low-energy constants (LECs) parametrizing chiral interactions. Assuming a simple model for these input uncertainties, we draw LECs from Gaussian distributions centered around the values of the NNLO$_{\textrm{opt}}$ parametrization \cite{Ekstrom13} with assumed standard deviations of $1\%$ of the central values. We find under these conditions $Q_2$ and $B(E2)$ distributions with standard deviations ranging from $3$-$6\%$ of the output mean values. Further, the variances in the \SpR{3} shape decompositions of the sampled nuclear states are remarkably small, while correlations describing the surface vibrations of these shapes vary appreciably more. Above all, we take steps toward a rigorous uncertainty analysis of \textit{ab initio} nuclear collectivity, with the eventual aim of understanding, from first principles, collective, clustering, and reaction features in nuclei while accounting for all sources of model uncertainty.

\section{\label{methods}Theoretical Methods}

\subsection{Many-body framework}
The most general approach to nuclear structure describes the atomic nucleus as a quantum $A$-body system that takes as input only the few-nucleon interactions (referred to here as \textit{ab initio} modeling) consistent with the symmetries of quantum chromodynamics (QCD). We utilize the SA-NCSM \cite{DytrychSBDV_PRL07,LauneyDD16,DytrychLDRWRBB20,LauneyMD_ARNPS21}, which is based on the NCSM concept \cite{NavratilVB00,BarrettNV13}, to solve the $A$-body Schrödinger equation for low-lying energy eigenstates of the nuclear Hamiltonian. Both the NCSM and the SA-NCSM make use of many-body basis states that are constructed from single-particle 3D harmonic oscillator (HO) wave functions, which are analytically known and easy to work with. It is natural to organize the many-body basis states in increasing HO energy, and to cut the infinitely-large basis at some maximum number of HO excitations ($N_{\textrm{max}}$). 

This truncation introduces uncertainty in all no-core shell model predictions, and while it is unavoidable it is controllable, i.e., this error systematically decreases as more HO shells are included in the calculation. These calculations additionally depend on the oscillator energy, \hw. This parameter can be understood as follows: larger \hw ~values lead to a better resolution, and to a smaller size of the box in which the nucleus resides. The HO cutoff $N_{\textrm{max}}$ can also be understood to limit the size of this box, which grows as more shells are included in the calculation. It is therefore important to consider a range of \hw ~and $N_{\textrm{max}}$ values, and to determine the convergence behavior of relevant observables when reporting \textit{ab initio} results. 

For a given inter-nucleon potential and considering all basis states up to $N_{\textrm{max}}$, the SA-NCSM yields exactly the same results as the NCSM. The key breakthrough of the SA-NCSM is that it further organizes the HO basis states into symmetry-adapted (SA) subspaces, allowing one to additionally select all necessary subspaces in a prescribed way (detailed in Ref. \cite{LauneyDSBD20}). This is done according to the algebraic properties of the deformation-preserving \SU{3} symmetry or the shape-preserving \SpR{3} symmetry which manifest in nuclei across the chart \cite{DytrychLDRWRBB20,BahriR00}. Even though the framework uses symmetry groups to construct and organize the basis states, calculations are not limited \textit{a priori} by any symmetry, and if the nuclear Hamiltonian demands, the method readily accommodates significant symmetry-breaking.

This study makes use of an \SpR{3}-adapted scheme, meaning that the many-body basis states are efficiently built using group theoretical algorithms according to the reduction chain $\SpR{3} \supset \SU{3}$. Each basis state is thus labelled by the \SpR{3} shape quantum numbers $N_{\sigma}(\lambda \, \mu)_{\sigma}$, the \SU{3} deformation quantum numbers $N_{\omega}(\lambda \, \mu)_{\omega}$, the total orbital angular momentum $L$, and the many-body spin configuration \{$S_p$, $S_n$, $S$\}. The \SU{3} numbers describe the spatial distribution of HO quanta for a many-body basis state, with $N_{\omega} = N_x + N_y + N_z$ indicating the total HO quanta above the valence configuration, $\lambda_{\omega} = N_z - N_x,$ and $\mu_{\omega} = N_x - N_y$. Hence, the SA-NCSM organizes the basis into shape-preserving subspaces (irreps) respecting \SpR{3} symmetry, each of which is spanned by many states all sharing the shape quantum numbers $N_{\sigma}(\lambda \, \mu)_{\sigma}$, but with differing deformation, understood as follows. The lowest HO-energy configuration is called the bandhead or equilibrium shape, and its deformation is given by the shape label: $N_{\sigma}(\lambda \, \mu)_{\sigma} = N_{\omega}(\lambda \, \mu)_{\omega}$. The remaining states are given by all of the parity-preserving particle-hole excitations of the giant-resonance type of the bandhead which preserve \SpR{3} symmetry, up to $N_{\textrm{max}}$ excitation quanta. They are interpreted as the surface vibrations which can be made on top of the bandhead and which preserve the equilibrium shape. 

\begin{figure*}
\centering
\includegraphics[width=\linewidth]{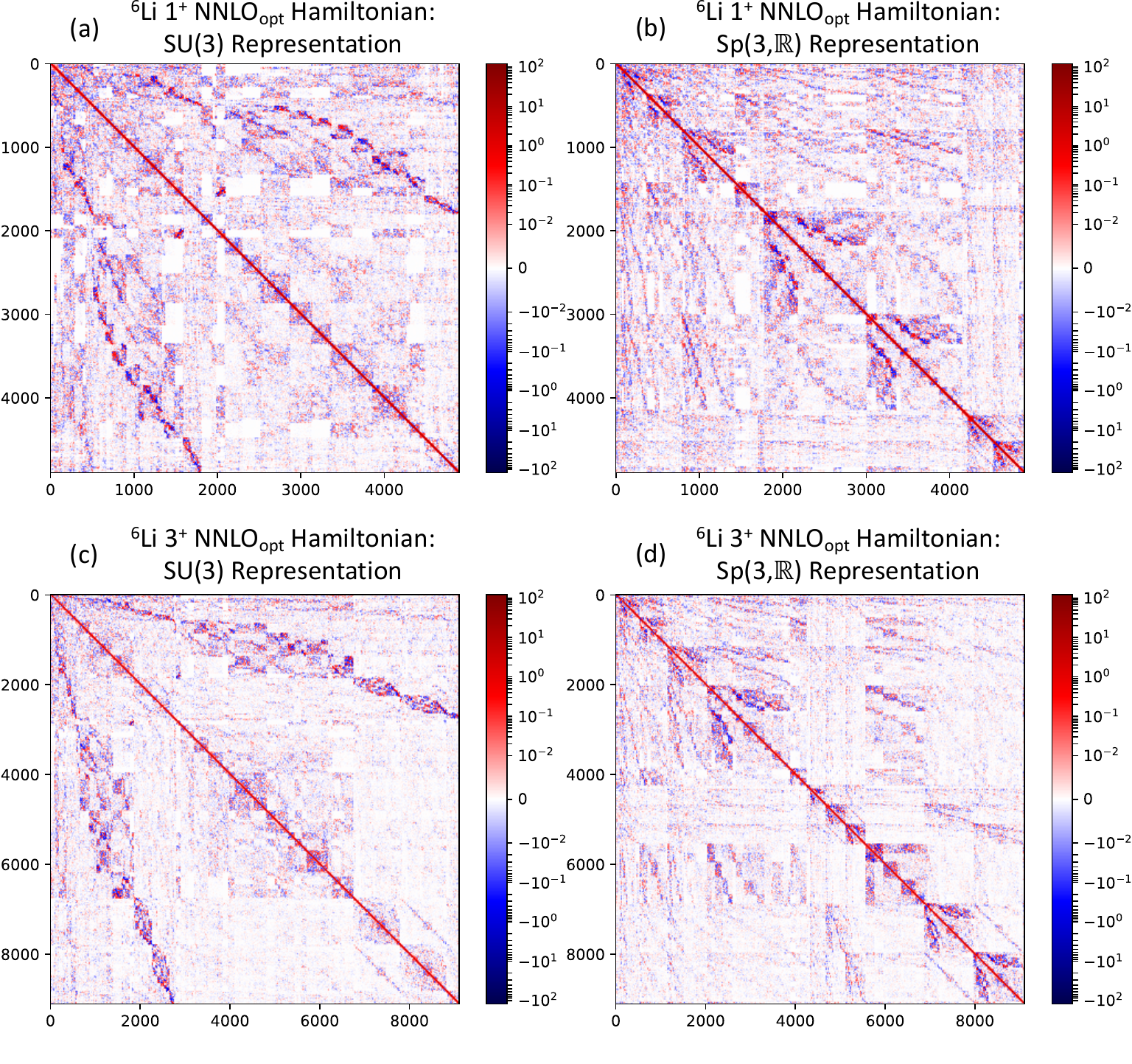}
\caption{Graphical representation of the NNLO$_{\textrm{opt}}$ Hamiltonian matrix elements calculated in the SA model spaces described in the text, including the kinetic energy and Coulomb interaction, computed for (a) \& (b) the $^6$Li $1^+$ ground state and (c) \& (d) the $^6$Li first excited $3^+$ state. The matrix elements are shown in two representations: for (a) \& (c), the bases are arranged into blocks of definite \SU{3} deformation, while for (b) \& (d) the bases are arranged into blocks of definite \SpR{3} shape. The numerical values of the matrix elements are mapped according to the color bars shown in a logarithmic scale.}
\label{6li_sp3r_MEs}
\end{figure*}

We find that realistic Hamiltonian matrices computed for the ground state $1^+$ and first excited $3^+$ state of $^6$Li with the SA-NCSM demonstrate remarkably ordered structure. When organized into \SU{3} subspaces and sorted according to increasing HO excitations $N_{\omega}$, the Hamiltonian matrix elements with the largest magnitudes lie along the main diagonal, but significant off-diagonal couplings are also present [FIG. \ref{6li_sp3r_MEs} (a) \& (c)]. These matrix elements describe couplings between low- and high-energy configurations of the nucleus, and are dispersed throughout the matrix. However, when organized explicitly into \SpR{3}-subspaces, sorted by decreasing probability amplitudes of the eigenfunctions of NNLO$_{\textrm{opt}}$, the matrices exhibit a clear block-diagonal structure [FIG. \ref{6li_sp3r_MEs} (b) \& (d)]. The off-diagonal couplings are much weaker, and indeed almost all of the largest matrix elements are grouped into blocks aggregated towards the main diagonal. This speaks to the power of the symplectic scheme. Even though both representations are calculated from the same inter-nucleon potential, organizing configurations according to their shape provides a block-diagonal matrix with the largest entries compressed along the diagonal, greatly reducing the computational load of solving for the low-energy eigenstates.

\textit{Ab initio} SA-NCSM calculations of low-lying states of light- to intermediate-mass nuclei reveal a striking and ubiquitous pattern \cite{DytrychLDRWRBB20,dytrychlmcdvl_prl12,DreyfussLTDBDB16,TobinFLDDB14,LauneyDSBD20}. Out of the many shapes which span a typical no-core shell-model space, only a very small subset account for the entire nuclear state, with one or two shapes tending to play a highly dominant role. Because the monopole operator $r^2$ and the quadrupole operator $Q_2$ do not mix nuclear shapes, collectivity is driven overwhelmingly by the few shapes that contribute at an appreciable level to the nuclear state, rendering symplectic symmetry a critically important symmetry of low-energy nuclear physics (for more details, refer to \cite{LauneyDD16}). 

\subsection{Inter-nucleon interaction}
Within this many-body approach, we utilize realistic nucleon-nucleon interactions based on chiral effective field theory \cite{Machleidt_2011}. These NN potentials express all of the interactions between the low-energy degrees of freedom (nucleons and pions) which conform to the symmetries of QCD, as an infinite expansion which must be truncated at some order in the perturbation theory. The derived forces at a given order are characterized by a number of unknown coefficients known as low-energy constants (LECS) which must be fit to few-nucleon experimental data. The LECs parametrize the unresolved low-energy quark-gluon dynamics, and provide a link between the effective theory and the underlying physics of QCD. The predictive power of \textit{ab initio} nuclear modeling is therefore highly dependent on the probability distribution of the fitted values of the LECs, which is a source of uncertainties that propagates into the many-body observables. Indeed, one must account for other sources of uncertainties, such as the momentum cutoff and regularization \cite{Gasparyan_2023}, and the truncation error at a given chiral order \cite{Furnstahl_2015}, among other considerations \cite{svensson2023inference,Chang_2018}.

Specifically, in addition to the long-ranged attractive one-pion exchange potential, at leading order (LO) in the expansion two short-ranged and repulsive NN contact forces emerge in $S$-waves, which yields four LECs in this study  (we note that for the $^1S_0$ partial wave we use an LEC for each of the three isospin triplet components). At next-to-leading order (NLO), seven repulsive contact terms in $S$- and $P$-waves enter with seven new LECs, as well as leading contributions to the two-pion exchange interaction. Important sub-leading corrections to the two-pion exchange are introduced at next-to-next-to-leading order (NNLO), bringing in three LECs for a total of fourteen up through NNLO in the two-nucleon sector. Three-nucleon forces also enter at NNLO, but for the purposes of this study they are neglected. These fourteen LECs are traditionally fit to few-nucleon scattering data while critical observables of light nuclei are also monitored. By using chiral potentials within the framework of the \textit{ab initio} SA-NCSM, in this study we quantify the uncertainties of nuclear collective observables that arise from variations in the LECs of the chiral potentials.

\section{\label{results}Results and Discussion}

We quantify how uncertainties in the chiral NNLO LECs translate into uncertainties on electric quadrupole $Q_2$ moments and $B(E2)$ transition rates, and further study the impact of these uncertainties on critical correlations in the nuclear states. Using \SpR{3} bases, we construct SA model spaces for the $^6$Li $1^+$ ground state, its first excited $3^+$ state, as well as the $0^+$ ground state of $^{12}$C and its first excited $2^+$ state. For the $^6$Li states, we consider 13 of the most important shapes as determined by complete $N_{\textrm{max}} = 12$ SA-NCSM calculations, and extend them to $N_{\textrm{max}} = 8$. We additionally include all \SU{3} configurations labelled by $N_{\omega} = 0$ and $N_{\omega} = 2$ HO excitations above the valence configuration. For the $^{12}$C states, we consider all 12 shapes that start at $N_{\sigma} = 0$, in addition to the 2(6 2) and 4(12 0) \SpR{3} irreps which are known to dominate the dynamics of the low-lying $0^+$ states \cite{DreyfussLTDB13}, and allow all of these shapes to develop up to $N_{\textrm{max}} = 6$. 

We calculate the matrix elements of each LEC-dependent part of the NNLO interaction separately in each model space, using an oscillator energy \hw ~$=15$ MeV (while here, for error estimates, we use an optimal \hw~value, that is, a shell spacing that provides a fast convergence of nuclear radii, different \hw~values could and should be considered for robust uncertainty quantification). In this way, an arbitrary parametrization is obtained by multiplying each term by the corresponding LEC value and summing the result into one matrix. The complete Hamiltonian is obtained by including all LEC-independent chiral interactions as well as the kinetic energy and Coulomb interaction. For all of our calculations, we utilize $300$ LECs sets sampled around a successful chiral potential. We use the NNLO$_{\rm opt}$ parametrization, which has been found to reproduce various observables without three-nucleon forces, including the $^4$He electric dipole polarizability \cite{BakerLBND20}; the challenging analyzing power for elastic proton scattering on $^4$He, $^{12}$C, and $^{16}$O \cite{BurrowsEWLMNP19}; along with $B(E2)$ transition strengths for $^{21}$Mg and $^{21}$F \cite{Ruotsalainen19} in the SA-NCSM without effective charges; as well as reaction observables \cite{Miller22}. As such we draw uncorrelated LEC samples in the vicinity of NNLO$_{\rm opt}$, and begin our discussion with this successful parametrization.

For each of the 14 LECs at NNLO, we construct a Gaussian distribution whose mean is the NNLO$_{\textrm{opt}}$ value of that LEC, and whose standard deviation is $1\%$ of that value. We draw $300$ samples from each of these distributions to obtain $300$ LECs sets to model an approximate input standard deviation of $1\%$ on each of our interaction parameters. We then use these parameter sets to calculate the nuclear states, $Q_2$ moments, and select $B(E2)$ transitions of the four states of interest. It is important to note that we consider only uncorrelated uncertainties in the values of the LECs, at present neglecting for example uncertainty in the chiral expansion parameter and the error in truncating the chiral expansion. At this stage we limit our analysis to the statistical properties of the output distributions given the assumed Gaussian distributions and variances on the LECs.

\begin{figure*}
\centering
\includegraphics[width=\linewidth]{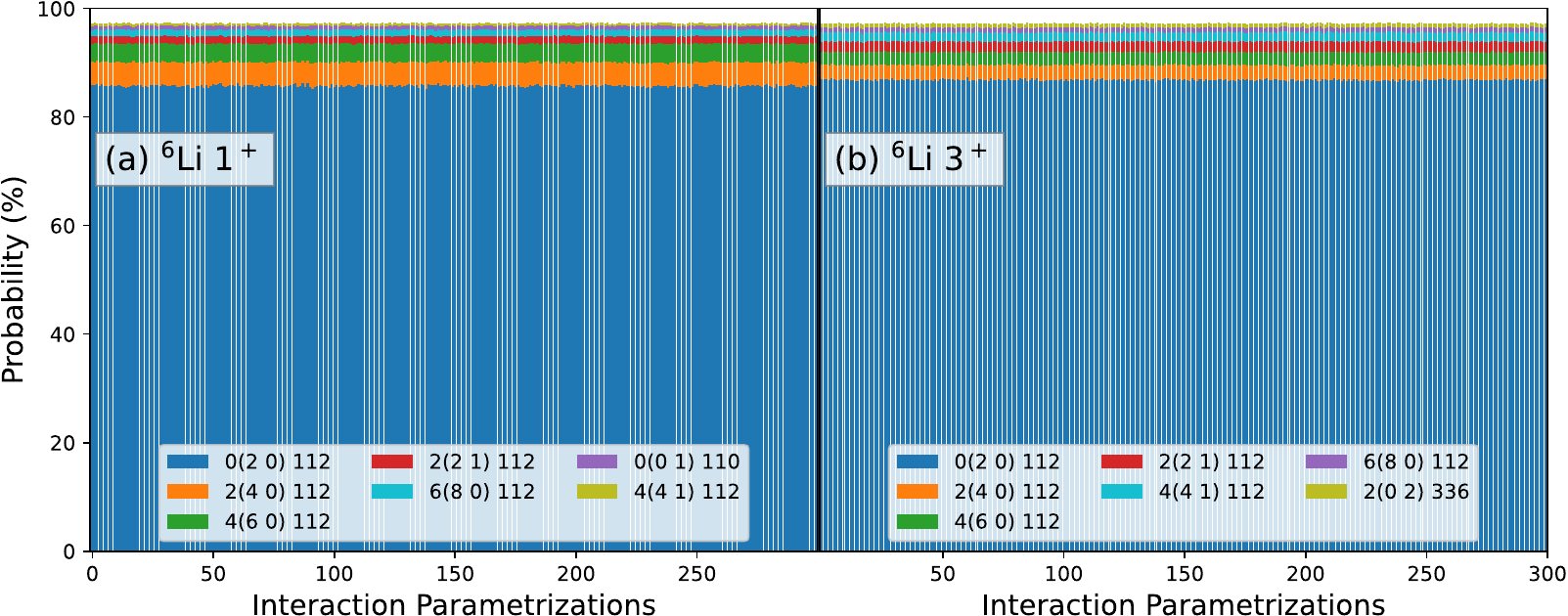}
\caption{Probability amplitudes of the most dominant shapes, labeled by the $N_{\sigma}(\lambda \, \mu)_{\sigma} 2S_p 2S_n 2S$ quantum numbers, present in (a) the $^6$Li $1^+$ ground state and (b) the $^6$Li first excited $3^+$ state, in $N_{\textrm{max}} = 8$ SA model spaces spanned by 13 shapes, obtained from $300$ sets of LECs simultaneously sampled and drawn from Gaussian distributions centered around the NNLO$_{\textrm{opt}}$ value and standard deviations taken as $1\%$ of this central value.}
\label{6li_sp3r_wfns}
\end{figure*}

\begin{figure*}
\centering
\includegraphics[width=\linewidth]{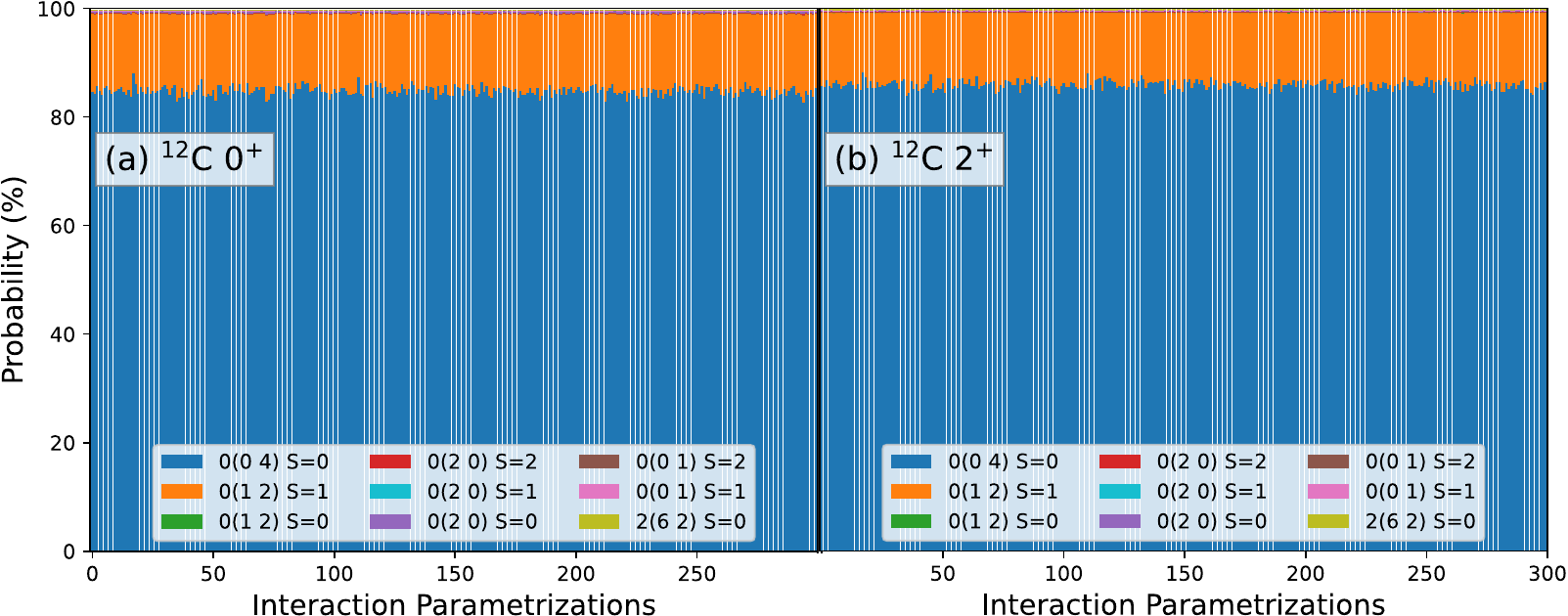}
\caption{Probability amplitudes of the most dominant shapes, labeled by the $N_{\sigma}(\lambda \, \mu)_{\sigma} S$ quantum numbers, present in (a) the $^{12}$C $0^+$ ground state and (b) the $^{12}$C first excited $2^+$ state, in $N_{\textrm{max}} = 6$ SA model spaces spanned by 14 shapes, obtained from the same LECs sets as FIG. \ref{6li_sp3r_wfns}.}
\label{12c_sp3r_wfns}
\end{figure*}

We find that the shape contributions within the $^6$Li $1^+$ and $3^+$ states remain almost constant with respect to our LECs samples, indicating that a $1\%$ uncertainty in the LEC values bears minimal impact on the symplectic symmetry manifesting in this nucleus (FIG. \ref{6li_sp3r_wfns}). In both states, the sampled nuclear states almost exactly reproduce those obtained with the NNLO$_{\textrm{opt}}$ LECs, with very small deviations. The dominant shape is always the prolate 0(2 0), followed secondarily by the increasingly prolate 2(4 0), 4(6 0), and 6(8 0) shapes. Similarly, the shape decompositions of the $^{12}$C ground state $0^+$ and first excited $2^+$ state vary minimally with the LECs, and the sampled nuclear states again very closely reproduce those of NNLO$_{\textrm{opt}}$ (FIG. \ref{12c_sp3r_wfns}). The dominant 0(0 4) shape is oblate, while two oblate 0(1 2) shapes with total spin = $1$ (the sum of which is considered) comprise most of the remainder of the states. The NNLO$_{\textrm{opt}}$ probabilities of the two most dominant shapes in all four states, along with the sampled means and standard deviations, are provided in TABLE \ref{sp3r_table}. We note that the shape contributions remain relatively constant even for a $50\%$ uniform variation of the LEC uncertainties \cite{Becker_CGS_2023}.

\begin{table*}[]
\begin{center}
\begin{tabular}{|l|l|l|l|l|l|l|l|l|l|}
\hline
Nucleus  & $J^{\pi}$ & \begin{tabular}[c]{@{}l@{}}Dominant\\ Shape\\$N_{\sigma}(\lambda \, \mu)_{\sigma} S$\end{tabular} & \begin{tabular}[c]{@{}l@{}}NNLO$_{\textrm{opt}}$\\ Probability\\ (\%)\end{tabular} & \begin{tabular}[c]{@{}l@{}}Mean\\ Probability\\ (\%)\end{tabular} & \begin{tabular}[c]{@{}l@{}}Standard\\ Deviation\\ (\%)\end{tabular} & \begin{tabular}[c]{@{}l@{}}Next\\ Dominant\\ Shape\end{tabular} & \begin{tabular}[c]{@{}l@{}}NNLO$_{\textrm{opt}}$\\ Probability\\ (\%)\end{tabular} & \begin{tabular}[c]{@{}l@{}}Mean\\ Probability\\ (\%)\end{tabular} & \begin{tabular}[c]{@{}l@{}}Standard\\ Deviation\\ (\%)\end{tabular} \\ \hline
$^6$Li   & $1^+$     & 0(2 0) $S = 1$                                           & $85.74$                                                                              & $85.75$                                                             & $0.19$                                                                & 2(4 0) $S = 1$                                                  & $4.35$                                                                               & $4.33$                                                              & $0.12$                                                                \\ \hline
$^6$Li   & $3^+$     & 0(2 0) $S = 1$                                           & $86.83$                                                                              & $86.84$                                                             & $0.14$                                                                & 2(4 0) $S = 1$                                                  & $2.73$                                                                               & $2.72$                                                              & $0.06$                                                                \\ \hline
$^{12}$C & $0^+$     & 0(0 4) $S = 0$                                           & $84.79$                                                                              & $84.83$                                                             & $0.86$                                                                & 0(1 2) $S = 1$                                                  & $14.20$                                                                              & $14.15$                                                             & $0.79$                                                                \\ \hline
$^{12}$C & $2^+$     & 0(0 4) $S = 0$                                           & $85.88$                                                                              & $85.89$                                                             & $0.79$                                                                & 0(1 2) $S = 1$                                                  & $13.30$                                                                              & $13.28$                                                             & $0.77$                                                                \\ \hline
\end{tabular}
\caption {Tabulated values of the two most dominant shapes [\SpR{3} bandheads] in the $^6$Li $1^+$ ground state, first excited $3^+$ state, $^{12}$C $0^+$ ground state, and first excited $2^+$ state sampled from the LEC sets described in the text. Reported are the probabilities obtained with NNLO$_{\textrm{opt}}$, as well as the sample distribution means and standard deviations.} \label{sp3r_table}  
\end{center}
\end{table*}

\begin{figure*}
\centering
\includegraphics[width=\linewidth]{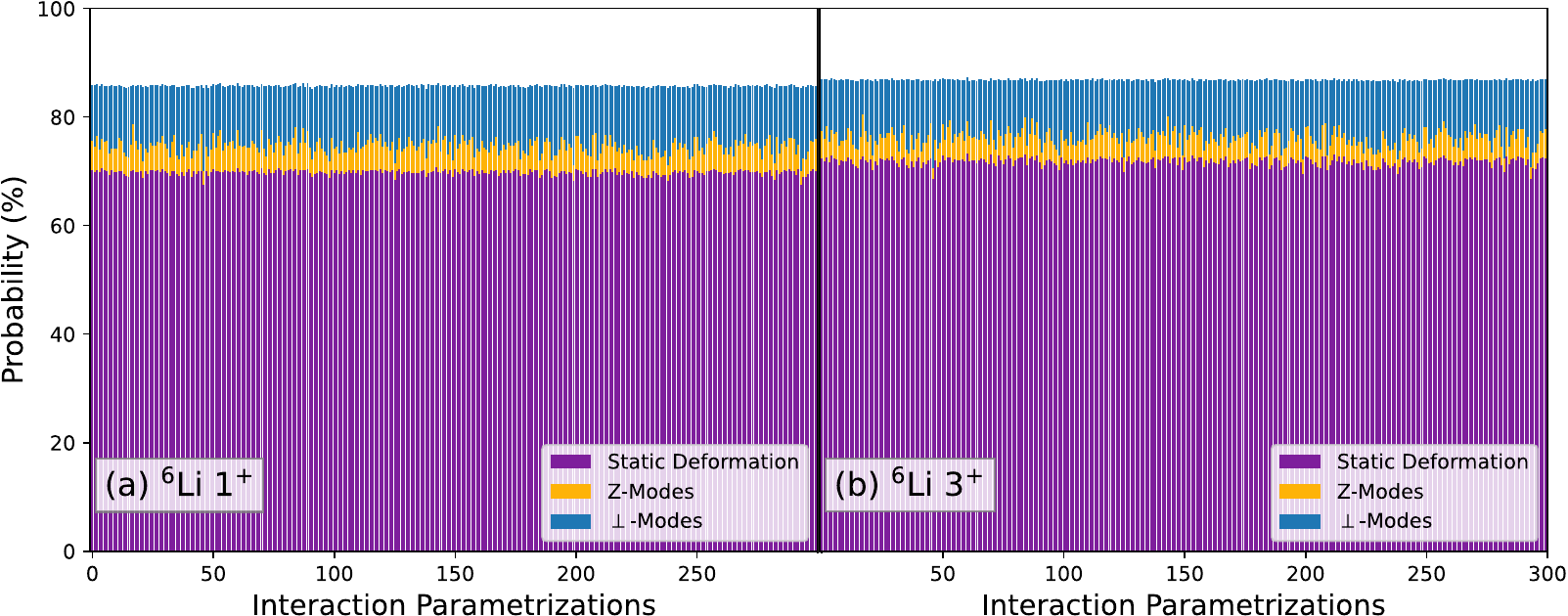}
\caption{The SU(3) breakdown of the dominant shape of (a) the $^6$Li $1^+$ ground state and (b) the $^6$Li first excited $3^+$ state, taken from the same sampled states as FIG. \ref{6li_sp3r_wfns}. Shown are the static deformation (purple), all surface vibrations in the $z$-direction (gold), and all remaining surface vibrations (blue).}
\label{6li_su3_wfns}
\end{figure*}

\begin{figure*}
\centering
\includegraphics[width=\linewidth]{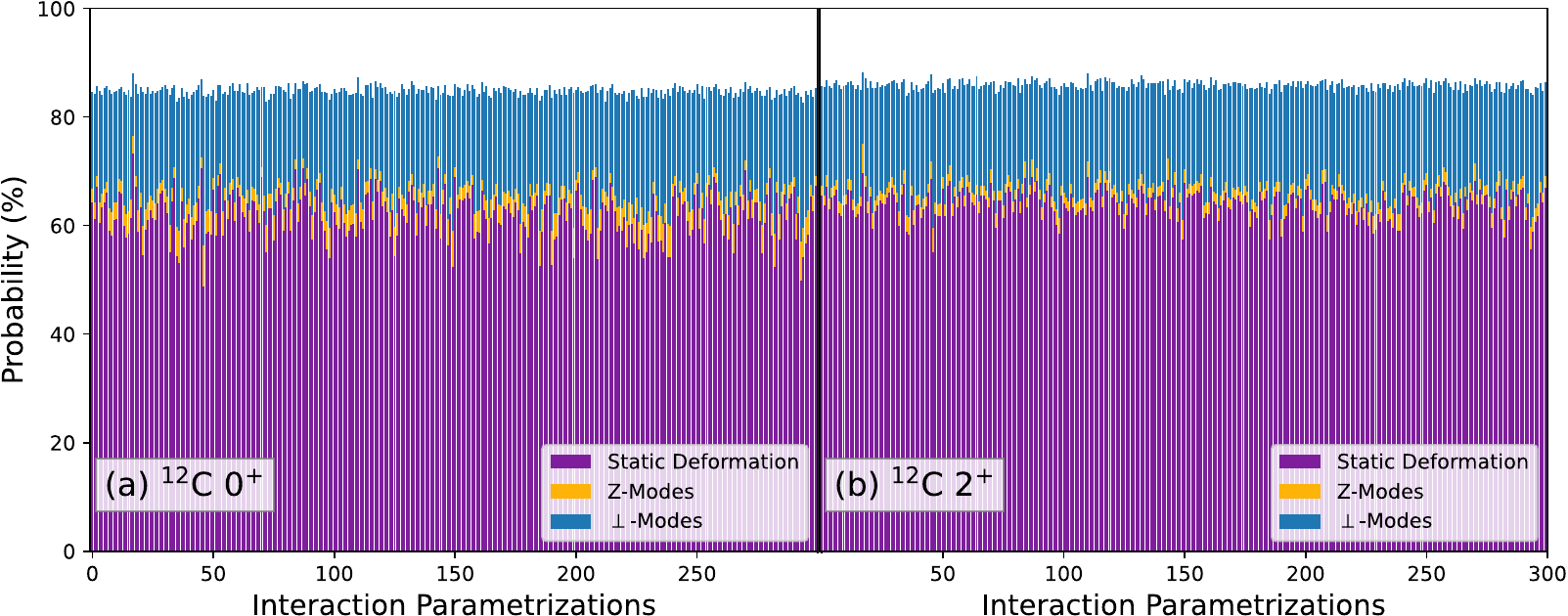}
\caption{The SU(3) breakdown of the dominant shape of (a) the $^{12}$C $0^+$ ground state and (b) the $^{12}$C first excited $2^+$ state, taken from the same sampled states as FIG. \ref{12c_sp3r_wfns}. Shown are the static deformation (purple), all surface vibrations in the $z$-direction (gold), and all remaining surface vibrations (blue).}
\label{12c_su3_wfns}
\end{figure*}

\begin{table*}[]
\begin{center}
\begin{tabular}{|l|l|l|l|l|l|l|l|}
\hline
Nucleus  & $J^{\pi}$ & \begin{tabular}[c]{@{}l@{}}Mean Static\\ Deformation \\ Probability (\%)\end{tabular} & \begin{tabular}[c]{@{}l@{}}Standard\\ Deviation (\%)\end{tabular} & \begin{tabular}[c]{@{}l@{}}Mean $Z$-Modes\\ Probability (\%)\end{tabular} & \begin{tabular}[c]{@{}l@{}}Standard\\ Deviation (\%)\end{tabular} & \begin{tabular}[c]{@{}l@{}}Mean $\perp$-Modes\\ Probability (\%)\end{tabular} & \begin{tabular}[c]{@{}l@{}}Standard\\ Deviation (\%)\end{tabular} \\ \hline
$^6$Li   & $1^+$     & $67.79$                                                                                 & $0.50$                                                              & $4.82$                                                                      & $1.06$                                                              & $11.14$                                                                         & $1.33$                                                              \\ \hline
$^6$Li   & $3^+$     & $71.79$                                                                                 & $0.77$                                                              & $4.32$                                                                      & $0.97$                                                              & $10.73$                                                                         & $1.59$                                                              \\ \hline
$^{12}$C & $0^+$     & $62.03$                                                                                 & $4.23$                                                              & $3.25$                                                                      & $1.16$                                                              & $19.54$                                                                         & $2.55$                                                              \\ \hline
$^{12}$C & $2^+$     & $64.05$                                                                                 & $2.60$                                                              & $2.10$                                                                      & $0.47$                                                              & $19.74$                                                                         & $1.87$                                                              \\ \hline
\end{tabular}
\caption {Tabulated values of the \SU{3} composition of the dominant shapes in the $^6$Li $1^+$ ground state, first excited $3^+$ state, $^{12}$C $0^+$ ground state, and first excited $2^+$ state sampled from the LEC sets described in the text. The probabilities are divided into three categories: the bandhead or static deformation, all \SpR{3} excitations which add quanta exclusively in the $z$-direction (``$z$-modes''), and all remaining \SpR{3} excitations (``perpendicular-modes''). Reported are the sample distribution means and standard deviations.} \label{su3_table} 
\end{center}
\end{table*}
\pagebreak
It is evident that symplectic symmetry is of critical importance to the states discussed, and that it is a very good symmetry of the NNLO Hamiltonian. The sub-percent standard deviations show that the assumed LEC uncertainties do not disrupt the most important features of the NNLO$_{\textrm{opt}}$ nuclear states. Not only are the dominant symplectic-preserving structures kept intact, but indeed almost all of the symplectic content is unaffected by the variance in the LEC samples. We can thus confidently assert that if uncertainties on the LECs are well-constrained for high-precision calculations, the most important symplectic features will also be well-constrained with errors on order of a few per cent at most, and ideally at the sub-percent level.

Although the symplectic decomposition of each low-lying nuclear state is unaffected, we observe slightly larger variations within the \SU{3} content of the dominant shapes themselves. We organize the \SU{3} configurations of these shapes into three categories: the bandhead basis state is referred to as that shape's static deformation; all symplectic excitations of the bandhead that add energy quanta exclusively in the $z$-direction are summed together and referred to as ``$z$-modes''; and all remaining symplectic excitations are summed together and referred to as ``perpendicular-modes''. 

The \SU{3} breakdowns of the 0(2 0) shape in the $^6$Li $1^+$ and $3^+$ are very similar (FIG. \ref{6li_su3_wfns}). The static deformation comprises most of the \SpR{3} irrep, followed by the perpendicular surface vibrations and lastly the $z$-modes. The largest standard deviations, both absolutely and as fractions of the corresponding output averages, arise in the perpendicular modes, while the smallest standard deviations come from the static deformation. The \SU{3} contents of the sampled $^{12}$C states tell a similar story, but with larger observed standard deviations compared to $^6$Li (FIG. \ref{12c_su3_wfns}). Again the most substantial contributions to the dominant 0(0 4) shape are the static deformation, followed by the perpendicular-modes and lastly by the $z$-modes. In these states however, it is the static deformation which exhibits the greatest standard deviation, and the $z$-modes vary the least. Further, all of the standard deviations quoted here are several times larger than the standard deviations observed in the total 0(2 0) and 0(0 4) shape probabilities, indicating more uncertainty in the composition of the dominant shape than in its overall probability. The reported means, standard deviations, and NNLO$_{\textrm{opt}}$ values are provided in TABLE \ref{su3_table}.

\begin{figure}
\centering
\includegraphics[width=\linewidth]{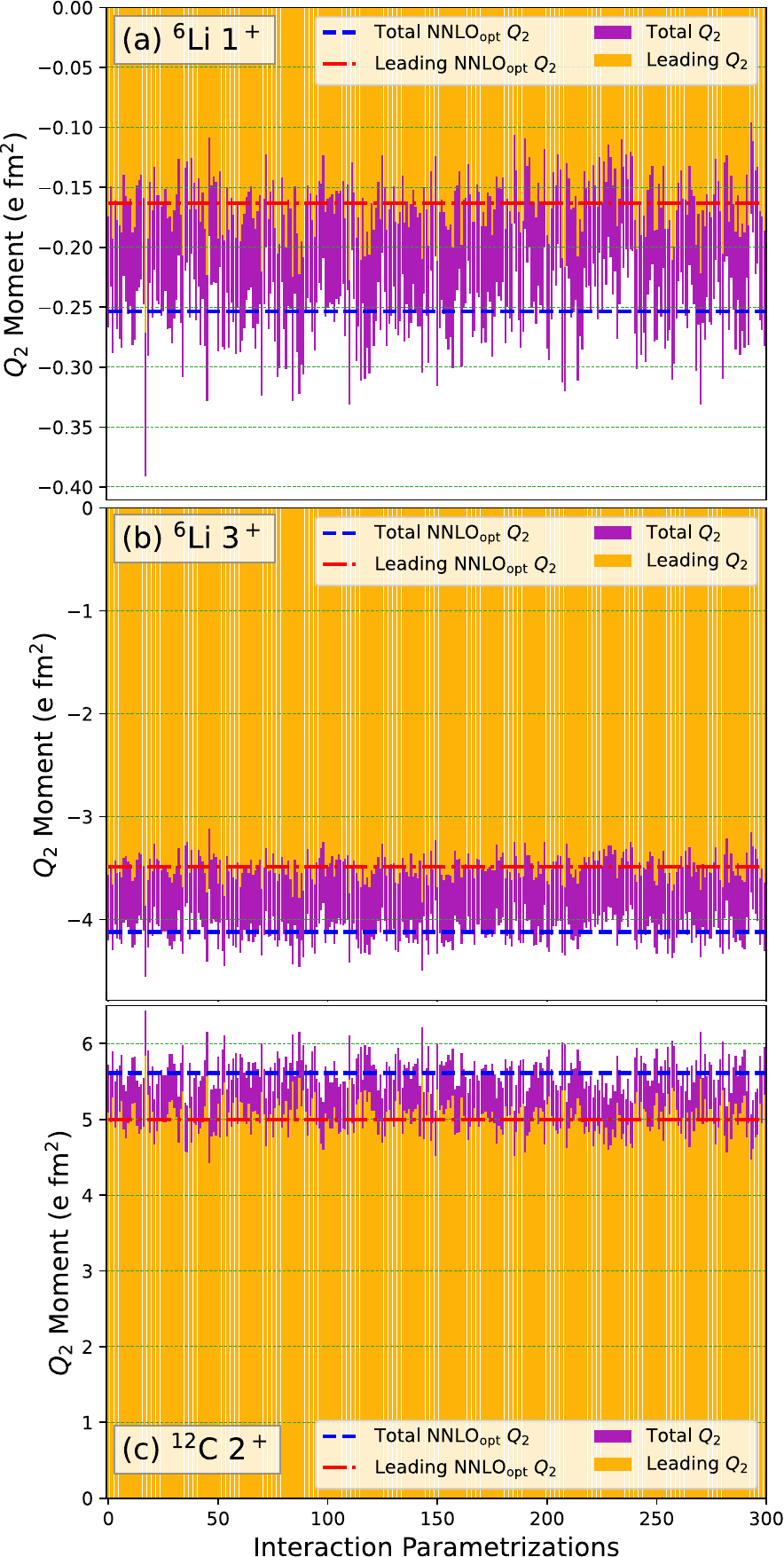}
\caption{The total $Q_2$ moments (purple) and the leading contributions to $Q_2$ from the dominant shape (gold), represented as histograms, obtained for (a) the $^6$Li $1^+$ ground state, (b) the $^6$Li first excited $3^+$ state, and (c) the $^{12}$C first excited $2^+$ state, from the sampled states discussed. The NNLO$_{\textrm{opt}}$ total $Q_2$ moments are indicated with blue dashed lines, while the NNLO$_{\textrm{opt}}$ leading $Q_2$ contributions are shown with red dashed-dotted lines.}
\label{leading_vs_total_Q2}
\end{figure}

\begin{table*}[]
\begin{center}
\begin{tabular}{|l|l|l|l|l|l|l|l|l|l|}
\hline
Nucleus  & $J^{\pi}$ & \begin{tabular}[c]{@{}l@{}}NNLO$_{\textrm{opt}}$\\ Total $Q_2$\\ (e fm$^2$)\end{tabular} & \begin{tabular}[c]{@{}l@{}}Mean\\ Total $Q_2$\\ (e fm$^2$)\end{tabular} & \begin{tabular}[c]{@{}l@{}}Standard\\ Deviation\\ (e fm$^2$)\end{tabular} & \begin{tabular}[c]{@{}l@{}}Standard\\ Deviation\\ (\% Mean)\end{tabular} & \begin{tabular}[c]{@{}l@{}}NNLO$_{\textrm{opt}}$\\ Leading $Q_2$\\ (e fm$^2$)\end{tabular} & \begin{tabular}[c]{@{}l@{}}Mean\\ Leading $Q_2$\\ (e fm$^2$)\end{tabular} & \begin{tabular}[c]{@{}l@{}}Standard\\ Deviation\\ (e fm$^2$)\end{tabular} & \begin{tabular}[c]{@{}l@{}}Standard\\ Deviation\\ (\% Mean)\end{tabular} \\ \hline
$^6$Li   & $1^+$     & $-0.253$                                                                                   & $-0.254$                                                                  & $0.032$                                                                     & $12.6$                                                                     & $-0.163$                                                                                     & $-0.164$                                                                    & $0.025$                                                                     & $15.2$                                                                     \\ \hline
$^6$Li   & $3^+$     & $-4.12$                                                                                    & $-4.13$                                                                   & $0.140$                                                                     & $3.39$                                                                     & $-3.485$                                                                                     & $-3.49$                                                                     & $0.125$                                                                     & $3.58$                                                                     \\ \hline
$^{12}$C & $2^+$     & $5.61$                                                                                     & $5.62$                                                                    & $0.228$                                                                     & $4.05$                                                                     & $5.00$                                                                                       & $5.01$                                                                      & $0.229$                                                                     & $4.57$                                                                     \\ \hline
\end{tabular}
\caption{Tabulated values of the $^6$Li $1^+$ ground state, the $^6$Li first excited $3^+$ state, and the $^{12}$C first excited $2^+$ state total and leading $Q_2$ moments obtained from the sampled states discussed in the text. Reported are the values obtained with the NNLO$_{\textrm{opt}}$ LECs, the means of the sampled $Q_2$ distributions, and the standard deviations in units of e fm$^2$ and as percentages of the means.} \label{Q2_table}
\end{center}
\end{table*}

From the $^6$Li $1^+$, $3^+$, and $^{12}$C $2^+$ sampled states, we compute two sets of electric quadrupole moments: the total $Q_2$ moment considering all shapes included in the model space, and the leading component of $Q_2$ considering the predominant nuclear shape only. In all three states, the leading contribution of $Q_2$ tracks well with the total $Q_2$ moment (FIG. \ref{leading_vs_total_Q2}). In general, the two sets mirror each other; when the total $Q_2$ increases, the leading component increases by a roughly similar amount. It seems evident that they are correlated. We quantify the standard deviations of these $Q_2$ distributions in TABLE \ref{Q2_table}, but as a general remark there is a greater spread with respect to the NNLO$_{\textrm{opt}}$ values in the $1^+$ (FIG. \ref{leading_vs_total_Q2}a) than in the $3^+$ (FIG. \ref{leading_vs_total_Q2}b) and $2^+$ (FIG. \ref{leading_vs_total_Q2}c). Additionally, the leading contribution as a fraction of the total is smallest in the $1^+$ state, notably more so than in the other two states. 

\begin{figure*}
\centering
\includegraphics[width=\linewidth]{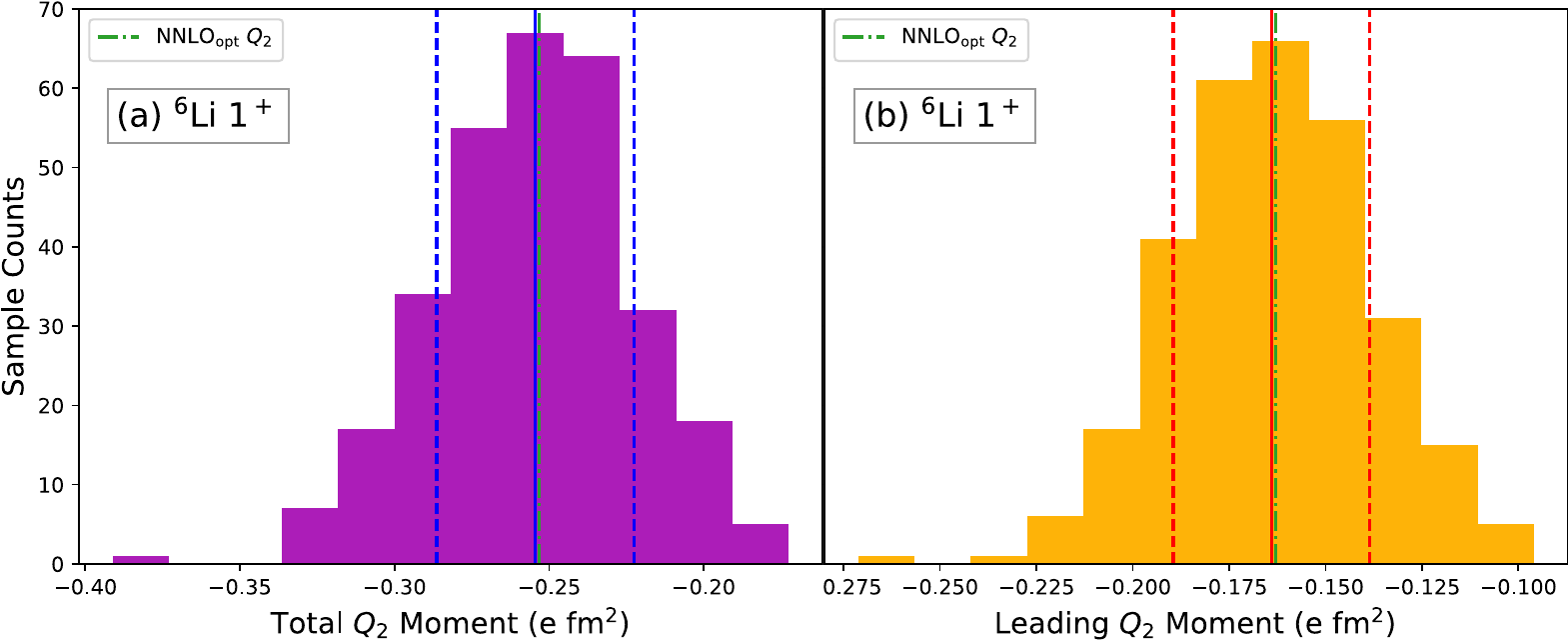}
\caption{Histograms of (a) the total $Q_2$ moments and (b) the leading contributions to the $Q_2$ moments sampled from the $^6$Li $1^+$ state discussed, arranged into 12 equally-spaced bins. The means of the distributions are indicated with a solid blue vertical line in (a) and a solid red vertical line in (b), with $1\sigma$ indicated by dashed blue lines in (a) and dashed red lines in (b). The values obtained with NNLO$_{\textrm{opt}}$ are indicated with dashed-dotted green lines.
}
\label{JJ2_hist}
\end{figure*}

We organize the sets of total and leading $Q_2$ moments into histograms divided into 12 equally-spaced bins, with the NNLO$_{\textrm{opt}}$ values as well as the distribution means and $1\sigma$ indicated. The total and leading $Q_2$ distributions of the $^6$Li $1^+$ state very closely resemble each other and share the same general features (FIG. \ref{JJ2_hist}). The observed standard deviations are not very large, but as percentages of their respective distribution means they are $12.6\%$ (a) and $15.2\%$ (b) -- quite large given the $1\%$ uncertainty in the sampled LECs. This might be a consequence of the very small quadrupole moment of the $1^+$ state of $^6$Li. The averages of both distributions are extremely close to the values obtained from the NNLO$_{\textrm{opt}}$ LECs, which is expected for Gaussian distributions with small variances; since our LEC distributions are centered around NNLO$_{\textrm{opt}}$, this is an important validation of our sampling. The averages and standard deviations are provided in TABLE \ref{Q2_table}.

\begin{figure*}
\centering
\includegraphics[width=\linewidth]{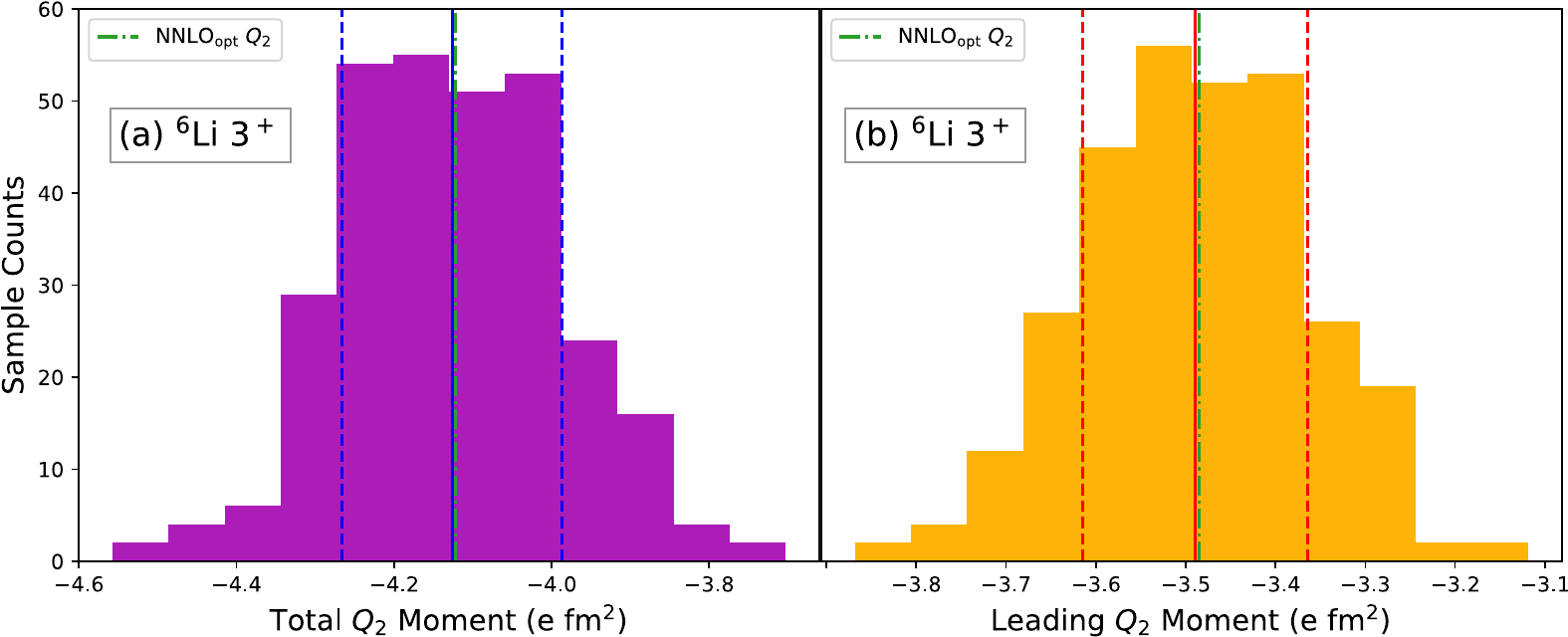}
\caption{
Histograms of (a) the total $Q_2$ moments and (b) the leading contributions to the $Q_2$ moments sampled from the $^6$Li $3^+$ state discussed, arranged into 12 equally-spaced bins. The means of the distributions are indicated with a solid blue vertical line in (a) and a solid red vertical line in (b), with $1\sigma$ indicated by dashed blue lines in (a) and dashed red lines in (b). The values obtained with NNLO$_{\textrm{opt}}$ are indicated with dashed-dotted green lines.
}
\label{JJ6_hist}
\end{figure*}

The distributions of the total and leading $Q_2$ moments are also very similar to each other in the $^6$Li $3^+$ state (FIG. \ref{JJ6_hist}). Their overall shapes are almost the same, and the reported standard deviations are very close to each other, suggesting that it is mostly the same physics driving the total $Q_2$ that is driving the contribution from the dominant shape. The averages and standard deviations are provided in TABLE \ref{Q2_table}. In e fm$^2$, the uncertainties are about an order of magnitude larger than those reported in the $1^+$, but are still rather small. More importantly, as percentages of the distribution means, the standard deviation of the total $Q_2$ is about $3.4\%$ and that of the leading piece is about $3.6\%$ -- better constrained than those of the $1^+$ by several factors. Clearly, if the LEC values can be well-constrained, so too can nuclear collectivity as expressed through both the symmetry content of the nuclear state as well as the quadrupole moment itself. We again note that the distribution means lie almost on top of the NNLO$_{\textrm{opt}}$ values.

\begin{figure*}
\centering
\includegraphics[width=\linewidth]{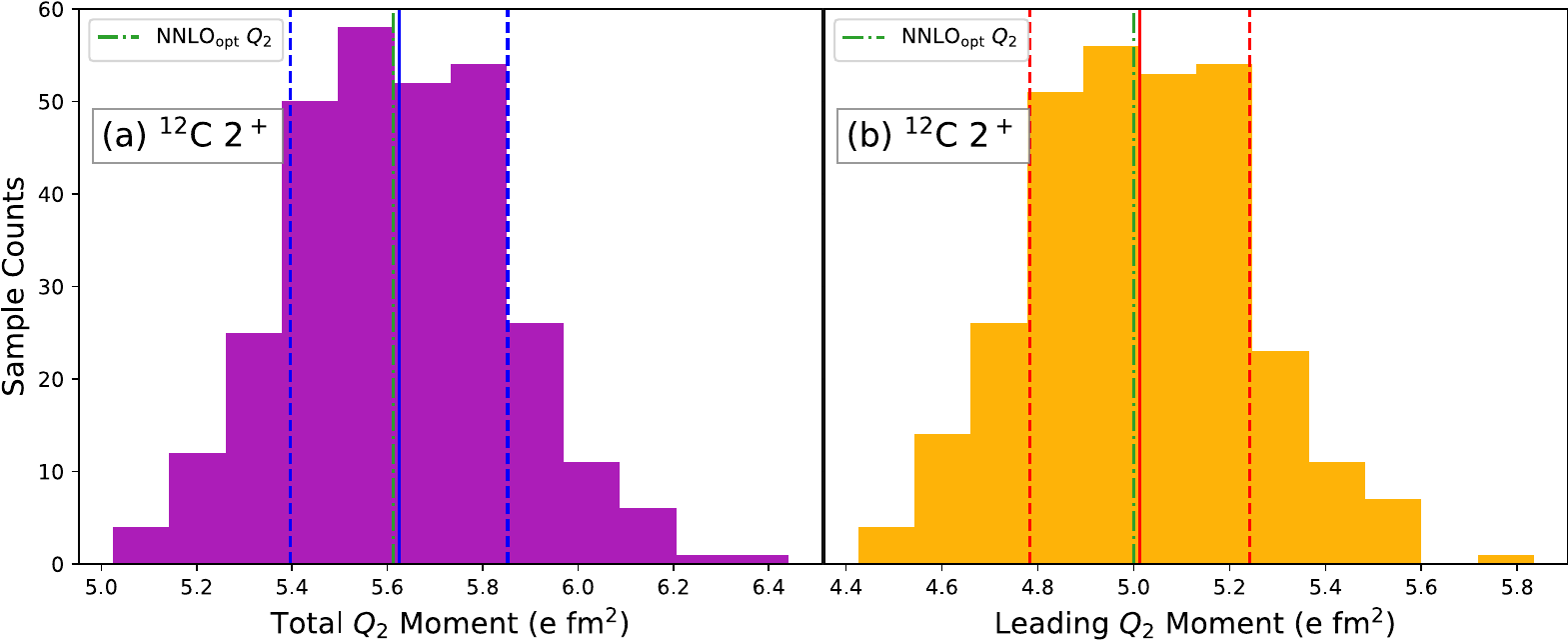}
\caption{Histograms of (a) the total $Q_2$ moments and (b) the leading contributions to the $Q_2$ moments sampled from the $^{12}$C $2^+$ state discussed, arranged into 12 equally-spaced bins. The means of the distributions are indicated with a solid blue vertical line in (a) and a solid red vertical line in (b), with $1\sigma$ indicated by dashed blue lines in (a) and dashed red lines in (b). The values obtained with NNLO$_{\textrm{opt}}$ are indicated with dashed-dotted green lines.}
\label{12C_hist}
\end{figure*}

The $Q_2$ moments of the $^{12}$C $2^+$ state tell a similar story to those of the $^6$Li $3^+$ (FIG. \ref{12C_hist}), suggesting their collectivity is very similar in nature. In e fm$^2$, the observed uncertainties in the $2^+$ are still fairly small, though they are about twice as large as those reported for the $3^+$. As percentages of the distribution means, the standard deviations are about $4.1\%$ for the total $Q_2$ and $4.6\%$ for the leading $Q_2$, only slightly larger than those of the $3^+$. Again the NNLO$_{\textrm{opt}}$ values lie extremely close to the distribution averages, and the shapes of the two distributions are very similar. This, and that the leading and total $Q_2$ variances are so close to each other, suggests that the same physics is driving both the total and leading component of the $2^+$ quadrupole moment. We additionally note that, although there is a clear preference to the 0(0 4), it does not dominate at such a high percentage as the 0(2 0) does in $^6$Li. This suggests that the 0(1 2) configurations of $^{12}$C contribute secondarily but not insignificantly to the total $Q_2$, perhaps more so than the next most probable shapes in $^6$Li do to its overall $Q_2$ moment. This may partially explain the slightly larger variations as compared to the $3^+$. The averages and standard deviations are provided in TABLE \ref{Q2_table}.

\begin{figure}
\centering
\includegraphics[width=\linewidth]{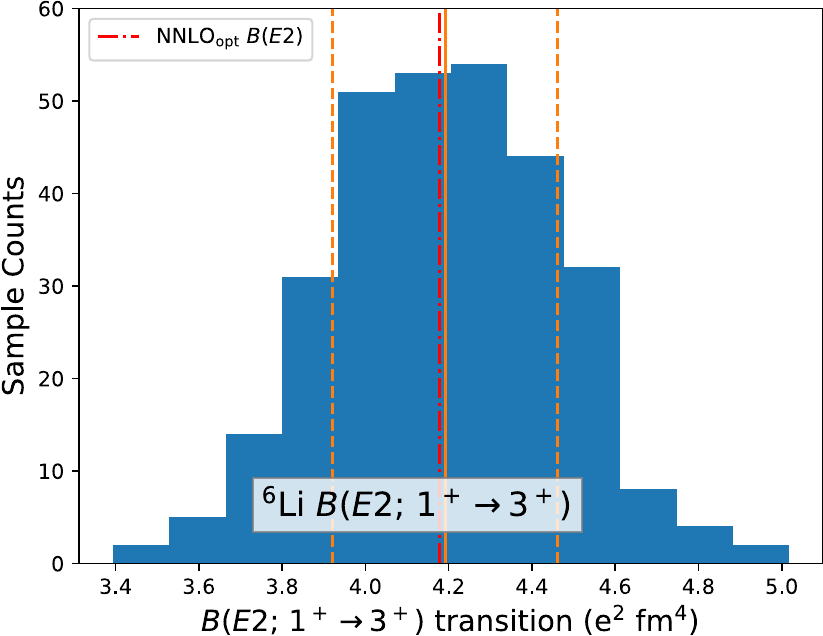}
\caption{Histogram of the sampled $B(E2; 1^+ \rightarrow 3^+)$ transition strengths of the sampled $^6$Li states discussed, arranged into 12 equally spaced bins. The distribution mean is indicated by a solid orange line, with $1\sigma$ indicated by dashed orange lines. The NNLO$_{\textrm{opt}}$ transition rate is indicated with a dashed-dotted red line.}
\label{E2_hist}
\end{figure}

Finally, we compute $B(E2; 1^+ \rightarrow 3^+)$ transition strengths for our $300$ $^6$Li states. The distribution is created from 12 equally spaced bins, and its shape looks like a good Gaussian distribution (FIG. \ref{E2_hist}). We find an average transition rate of $4.19$ e$^2$ fm$^4$, with a standard deviation of $0.27$ e$^2$ fm$^4$. As a percentage of the mean, this uncertainty is about $6.5\%$. We thus find that the $E2$ transition rates are reasonably well constrained, better than the $Q_2$ moments of the $1^+$ state but a little bit worse than the $Q_2$ moments of the $3^+$ state. Since the $E2$ transition takes in both the $1^+$ and $3^+$ states as input, uncertainty in the final output arises from both states, hence it lies between the uncertainties of the $1^+$ and $3^+$ quadrupole moments. The observed uncertainty in the $E2$ transition is more in line with the variance of the $3^+$ $Q_2$ moments rather than those of the $1^+$, because the $B(E2)$ value is dominated by the much larger $Q_2$ contribution of the $3^+$. We note that the distribution mean is again very close to the NNLO$_{\textrm{opt}}$ transition rate. 

In summary, we find that uncertainties on collective observables of the lowest-lying $^6$Li and $^{12}$C states are reasonably, but non-trivially, constrained assuming rather small uncertainties in the LECs parametrizing chiral nucleon-nucleon interactions at NNLO. Modeling these input uncertainties by drawing LEC samples from Gaussian distributions with a standard deviation of $1\%$ of the NNLO$_{\textrm{opt}}$ values, we find uncertainties as percentages of the mean quadrupole moment on order of $12\%$ in the $^6$Li $1^+$ ground state, $3\%$ in its first excited $3^+$ state, and $4\%$ in the first excited $2^+$ state of $^{12}$C. We similarly find an uncertainty on order of $6\%$ of the distribution mean for the $B(E2; 1^+ \rightarrow 3^+)$ transition rate of $^6$Li. The results suggest even at this level of input uncertainty, there is non-trivial uncertainty of at least a few per cent in the collective observables calculated. If the LEC values are less certain than our assumption, we can expect to see uncertainties greater than a few percent (perhaps significantly higher) in our predictions, and conversely if sub-percent accuracy is desired, the LEC values should be better constrained beyond $1\%$. We note that this conclusion is based only on a small variation of the LECs in the vicinity of NNLO$_{\rm opt}$.

Although our focus lies rightly with the observable properties of nuclei, it is also interesting to examine the variances of important structures in the nuclear states. Of particular note is that $3$-$12\%$ uncertainty is reported in the observables despite the fact that the symplectic composition of the states is almost unaffected by the choice of LECs. This indicates that a comparatively wide range of observables is realized with almost identical nuclear states – the dominant shapes always exhibit a standard deviation less than $1\%$. Clearly it is not enough to only look at the distribution of shapes in the state. It appears that the observable uncertainty is driven more by the internal composition of the dominant shape than it is by that shape's overall probability. This is in line with recent findings \cite{Becker_CGS_2023,Becker_PRL_2024}, though it is still unclear which aspects of this internal composition provide the greatest sensitivity to collective features.

Finally, it is important to note that this study is the first step towards a rigorous uncertainty quantification of \textit{ab initio} nuclear collectivity. Before drawing more robust conclusions, additional sources of uncertainty should be explicitly and simultaneously considered, such as uncertainties in the chiral EFT truncation error, from neglecting $3N$-forces, from choosing an oscillator energy for the basis, and so on. Although the SA model spaces are well-selected, they are developed only to relatively small $N_{\textrm{max}}$'s, so a full analysis would further benefit from expanded model spaces and a quantification of errors arising from the basis selection and $N_{\textrm{max}}$ cutoff, with the disadvantage of exponentially-growing computational costs. As a first step, however, we demonstrate an interesting array of features that further points us to a deeper understanding of the relationship between the nuclear interaction and emergent collectivity. We additionally lay the path forward to a rigorous uncertainty quantification on such collective observables, and towards the design of highly-optimized chiral potentials capable of precisely capturing challenging emergent structures and features across the nuclear chart. Such a rigorous uncertainty quantification entails achieving predictive LEC probability distributions, taking into account EFT truncation errors, SA-NCSM many-body uncertainties (see, e.g., \cite{Sargsyan_A8}), and utilizing the computational advantages of reduced basis methods like using the symplectic basis or emulation techniques (see, e.g., \cite{BeckerLED22}).

\begin{acknowledgments}
This work was supported by the U.S. National Science Foundation (PHY-2209060), as well as in part by the U.S. Department of Energy (DE-SC0023532), the European Research Council (ERC) under the European Union's Horizon 2020 research and innovation program (Grant Agreement No. 758027), the Swedish Research Council (Grant Agreement No. 2020- 05127), and the Czech Science Foundation (22-14497S). This material is based upon work supported by the U.S. Department of Energy, Office of Science, Office of Nuclear Physics, under the FRIB Theory Alliance award DE-SC0013617. This work benefited from high performance computational resources provided by LSU (www.hpc.lsu.edu), the National Energy Research Scientific Computing Center (NERSC), a U.S. Department of Energy Office of Science User Facility at Lawrence Berkeley National Laboratory operated under Contract No. DE-AC02-05CH11231, as well as the Frontera computing project at the Texas Advanced Computing Center, made possible by National Science Foundation award OAC-1818253.
\end{acknowledgments}

\bibliography{final_bib}

\begin{thebibliography}{35}%
\makeatletter
\providecommand \@ifxundefined [1]{%
 \@ifx{#1\undefined}
}%
\providecommand \@ifnum [1]{%
 \ifnum #1\expandafter \@firstoftwo
 \else \expandafter \@secondoftwo
 \fi
}%
\providecommand \@ifx [1]{%
 \ifx #1\expandafter \@firstoftwo
 \else \expandafter \@secondoftwo
 \fi
}%
\providecommand \natexlab [1]{#1}%
\providecommand \enquote  [1]{``#1''}%
\providecommand \bibnamefont  [1]{#1}%
\providecommand \bibfnamefont [1]{#1}%
\providecommand \citenamefont [1]{#1}%
\providecommand \href@noop [0]{\@secondoftwo}%
\providecommand \href [0]{\begingroup \@sanitize@url \@href}%
\providecommand \@href[1]{\@@startlink{#1}\@@href}%
\providecommand \@@href[1]{\endgroup#1\@@endlink}%
\providecommand \@sanitize@url [0]{\catcode `\\12\catcode `\$12\catcode `\&12\catcode `\#12\catcode `\^12\catcode `\_12\catcode `\%12\relax}%
\providecommand \@@startlink[1]{}%
\providecommand \@@endlink[0]{}%
\providecommand \url  [0]{\begingroup\@sanitize@url \@url }%
\providecommand \@url [1]{\endgroup\@href {#1}{\urlprefix }}%
\providecommand \urlprefix  [0]{URL }%
\providecommand \Eprint [0]{\href }%
\providecommand \doibase [0]{https://doi.org/}%
\providecommand \selectlanguage [0]{\@gobble}%
\providecommand \bibinfo  [0]{\@secondoftwo}%
\providecommand \bibfield  [0]{\@secondoftwo}%
\providecommand \translation [1]{[#1]}%
\providecommand \BibitemOpen [0]{}%
\providecommand \bibitemStop [0]{}%
\providecommand \bibitemNoStop [0]{.\EOS\space}%
\providecommand \EOS [0]{\spacefactor3000\relax}%
\providecommand \BibitemShut  [1]{\csname bibitem#1\endcsname}%
\let\auto@bib@innerbib\@empty
\bibitem [{\citenamefont {Wesolowski}\ \emph {et~al.}(2019)\citenamefont {Wesolowski}, \citenamefont {Furnstahl}, \citenamefont {Melendez},\ and\ \citenamefont {Phillips}}]{Wesolowski_2019}%
  \BibitemOpen
  \bibfield  {author} {\bibinfo {author} {\bibfnamefont {S.}~\bibnamefont {Wesolowski}}, \bibinfo {author} {\bibfnamefont {R.~J.}\ \bibnamefont {Furnstahl}}, \bibinfo {author} {\bibfnamefont {J.~A.}\ \bibnamefont {Melendez}},\ and\ \bibinfo {author} {\bibfnamefont {D.~R.}\ \bibnamefont {Phillips}},\ }\bibfield  {title} {\bibinfo {title} {Exploring bayesian parameter estimation for chiral effective field theory using nucleon–nucleon phase shifts},\ }\href {https://doi.org/10.1088/1361-6471/aaf5fc} {\bibfield  {journal} {\bibinfo  {journal} {Journal of Physics G: Nuclear and Particle Physics}\ }\textbf {\bibinfo {volume} {46}},\ \bibinfo {pages} {045102} (\bibinfo {year} {2019})}\BibitemShut {NoStop}%
\bibitem [{\citenamefont {Odell}\ \emph {et~al.}(2023)\citenamefont {Odell}, \citenamefont {Giuliani}, \citenamefont {Beyer}, \citenamefont {Catacora-Rios}, \citenamefont {Chan}, \citenamefont {Bonilla}, \citenamefont {Furnstahl}, \citenamefont {Godbey},\ and\ \citenamefont {Nunes}}]{odell2023rose}%
  \BibitemOpen
  \bibfield  {author} {\bibinfo {author} {\bibfnamefont {D.}~\bibnamefont {Odell}}, \bibinfo {author} {\bibfnamefont {P.}~\bibnamefont {Giuliani}}, \bibinfo {author} {\bibfnamefont {K.}~\bibnamefont {Beyer}}, \bibinfo {author} {\bibfnamefont {M.}~\bibnamefont {Catacora-Rios}}, \bibinfo {author} {\bibfnamefont {M.~Y.~H.}\ \bibnamefont {Chan}}, \bibinfo {author} {\bibfnamefont {E.}~\bibnamefont {Bonilla}}, \bibinfo {author} {\bibfnamefont {R.~J.}\ \bibnamefont {Furnstahl}}, \bibinfo {author} {\bibfnamefont {K.}~\bibnamefont {Godbey}},\ and\ \bibinfo {author} {\bibfnamefont {F.~M.}\ \bibnamefont {Nunes}},\ }\href@noop {} {\bibinfo {title} {Rose: A reduced-order scattering emulator for optical models}} (\bibinfo {year} {2023}),\ \Eprint {https://arxiv.org/abs/2312.12426} {arXiv:2312.12426 [physics.comp-ph]} \BibitemShut {NoStop}%
\bibitem [{\citenamefont {King}\ \emph {et~al.}(2019)\citenamefont {King}, \citenamefont {Lovell}, \citenamefont {Neufcourt},\ and\ \citenamefont {Nunes}}]{PhysRevLett.122.232502}%
  \BibitemOpen
  \bibfield  {author} {\bibinfo {author} {\bibfnamefont {G.~B.}\ \bibnamefont {King}}, \bibinfo {author} {\bibfnamefont {A.~E.}\ \bibnamefont {Lovell}}, \bibinfo {author} {\bibfnamefont {L.}~\bibnamefont {Neufcourt}},\ and\ \bibinfo {author} {\bibfnamefont {F.~M.}\ \bibnamefont {Nunes}},\ }\bibfield  {title} {\bibinfo {title} {Direct comparison between bayesian and frequentist uncertainty quantification for nuclear reactions},\ }\href {https://doi.org/10.1103/PhysRevLett.122.232502} {\bibfield  {journal} {\bibinfo  {journal} {Phys. Rev. Lett.}\ }\textbf {\bibinfo {volume} {122}},\ \bibinfo {pages} {232502} (\bibinfo {year} {2019})}\BibitemShut {NoStop}%
\bibitem [{\citenamefont {Hamaker}\ \emph {et~al.}(2021)\citenamefont {Hamaker}, \citenamefont {Leistenschneider}, \citenamefont {Jain}, , \citenamefont {Bollen}, \citenamefont {Giuliani}, \citenamefont {Lund}, \citenamefont {Nazarewicz}, \citenamefont {Neufcourt}, \citenamefont {Nicoloff}, \citenamefont {Puentes}, \citenamefont {Ringle}, \citenamefont {Sumithrarachchi},\ and\ \citenamefont {Yandow}}]{nat_bmm}%
  \BibitemOpen
  \bibfield  {author} {\bibinfo {author} {\bibfnamefont {A.}~\bibnamefont {Hamaker}}, \bibinfo {author} {\bibfnamefont {E.}~\bibnamefont {Leistenschneider}}, \bibinfo {author} {\bibfnamefont {R.}~\bibnamefont {Jain}}, , \bibinfo {author} {\bibfnamefont {G.}~\bibnamefont {Bollen}}, \bibinfo {author} {\bibfnamefont {S.~A.}\ \bibnamefont {Giuliani}}, \bibinfo {author} {\bibfnamefont {K.}~\bibnamefont {Lund}}, \bibinfo {author} {\bibfnamefont {W.}~\bibnamefont {Nazarewicz}}, \bibinfo {author} {\bibfnamefont {L.}~\bibnamefont {Neufcourt}}, \bibinfo {author} {\bibfnamefont {C.~R.}\ \bibnamefont {Nicoloff}}, \bibinfo {author} {\bibfnamefont {D.}~\bibnamefont {Puentes}}, \bibinfo {author} {\bibfnamefont {R.}~\bibnamefont {Ringle}}, \bibinfo {author} {\bibfnamefont {C.~S.}\ \bibnamefont {Sumithrarachchi}},\ and\ \bibinfo {author} {\bibfnamefont {I.~T.}\ \bibnamefont {Yandow}},\ }\bibfield  {title} {\bibinfo {title} {Precision mass measurement of lightweight self-conjugate nucleus 80zr},\ }\href
  {https://doi.org/10.1038/s41567-021-01395-w} {\bibfield  {journal} {\bibinfo  {journal} {Nat. Phys.}\ }\textbf {\bibinfo {volume} {17}},\ \bibinfo {pages} {1408} (\bibinfo {year} {2021})}\BibitemShut {NoStop}%
\bibitem [{\citenamefont {Wendt}\ \emph {et~al.}(2015)\citenamefont {Wendt}, \citenamefont {Forssén}, \citenamefont {Papenbrock},\ and\ \citenamefont {Sääf}}]{Wendt_2015}%
  \BibitemOpen
  \bibfield  {author} {\bibinfo {author} {\bibfnamefont {K.~A.}\ \bibnamefont {Wendt}}, \bibinfo {author} {\bibfnamefont {C.}~\bibnamefont {Forssén}}, \bibinfo {author} {\bibfnamefont {T.}~\bibnamefont {Papenbrock}},\ and\ \bibinfo {author} {\bibfnamefont {D.}~\bibnamefont {Sääf}},\ }\bibfield  {title} {\bibinfo {title} {Infrared length scale and extrapolations for the no-core shell model},\ }\bibfield  {journal} {\bibinfo  {journal} {Physical Review C}\ }\textbf {\bibinfo {volume} {91}},\ \href {https://doi.org/10.1103/physrevc.91.061301} {10.1103/physrevc.91.061301} (\bibinfo {year} {2015})\BibitemShut {NoStop}%
\bibitem [{\citenamefont {König}\ \emph {et~al.}(2014)\citenamefont {König}, \citenamefont {Bogner}, \citenamefont {Furnstahl}, \citenamefont {More},\ and\ \citenamefont {Papenbrock}}]{K_nig_2014}%
  \BibitemOpen
  \bibfield  {author} {\bibinfo {author} {\bibfnamefont {S.}~\bibnamefont {König}}, \bibinfo {author} {\bibfnamefont {S.~K.}\ \bibnamefont {Bogner}}, \bibinfo {author} {\bibfnamefont {R.~J.}\ \bibnamefont {Furnstahl}}, \bibinfo {author} {\bibfnamefont {S.~N.}\ \bibnamefont {More}},\ and\ \bibinfo {author} {\bibfnamefont {T.}~\bibnamefont {Papenbrock}},\ }\bibfield  {title} {\bibinfo {title} {Ultraviolet extrapolations in finite oscillator bases},\ }\bibfield  {journal} {\bibinfo  {journal} {Physical Review C}\ }\textbf {\bibinfo {volume} {90}},\ \href {https://doi.org/10.1103/physrevc.90.064007} {10.1103/physrevc.90.064007} (\bibinfo {year} {2014})\BibitemShut {NoStop}%
\bibitem [{\citenamefont {Coon}\ \emph {et~al.}(2012)\citenamefont {Coon}, \citenamefont {Avetian}, \citenamefont {Kruse}, \citenamefont {van Kolck}, \citenamefont {Maris},\ and\ \citenamefont {Vary}}]{PhysRevC.86.054002}%
  \BibitemOpen
  \bibfield  {author} {\bibinfo {author} {\bibfnamefont {S.~A.}\ \bibnamefont {Coon}}, \bibinfo {author} {\bibfnamefont {M.~I.}\ \bibnamefont {Avetian}}, \bibinfo {author} {\bibfnamefont {M.~K.~G.}\ \bibnamefont {Kruse}}, \bibinfo {author} {\bibfnamefont {U.}~\bibnamefont {van Kolck}}, \bibinfo {author} {\bibfnamefont {P.}~\bibnamefont {Maris}},\ and\ \bibinfo {author} {\bibfnamefont {J.~P.}\ \bibnamefont {Vary}},\ }\bibfield  {title} {\bibinfo {title} {Convergence properties of ab initio calculations of light nuclei in a harmonic oscillator basis},\ }\href {https://doi.org/10.1103/PhysRevC.86.054002} {\bibfield  {journal} {\bibinfo  {journal} {Phys. Rev. C}\ }\textbf {\bibinfo {volume} {86}},\ \bibinfo {pages} {054002} (\bibinfo {year} {2012})}\BibitemShut {NoStop}%
\bibitem [{\citenamefont {Furnstahl}\ \emph {et~al.}(2015)\citenamefont {Furnstahl}, \citenamefont {Klco}, \citenamefont {Phillips},\ and\ \citenamefont {Wesolowski}}]{Furnstahl_2015}%
  \BibitemOpen
  \bibfield  {author} {\bibinfo {author} {\bibfnamefont {R.~J.}\ \bibnamefont {Furnstahl}}, \bibinfo {author} {\bibfnamefont {N.}~\bibnamefont {Klco}}, \bibinfo {author} {\bibfnamefont {D.~R.}\ \bibnamefont {Phillips}},\ and\ \bibinfo {author} {\bibfnamefont {S.}~\bibnamefont {Wesolowski}},\ }\bibfield  {title} {\bibinfo {title} {Quantifying truncation errors in effective field theory},\ }\bibfield  {journal} {\bibinfo  {journal} {Physical Review C}\ }\textbf {\bibinfo {volume} {92}},\ \href {https://doi.org/10.1103/physrevc.92.024005} {10.1103/physrevc.92.024005} (\bibinfo {year} {2015})\BibitemShut {NoStop}%
\bibitem [{\citenamefont {Wesolowski}\ \emph {et~al.}(2021)\citenamefont {Wesolowski}, \citenamefont {Svensson}, \citenamefont {Ekstr\"om}, \citenamefont {Forss\'en}, \citenamefont {Furnstahl}, \citenamefont {Melendez},\ and\ \citenamefont {Phillips}}]{PhysRevC.104.064001}%
  \BibitemOpen
  \bibfield  {author} {\bibinfo {author} {\bibfnamefont {S.}~\bibnamefont {Wesolowski}}, \bibinfo {author} {\bibfnamefont {I.}~\bibnamefont {Svensson}}, \bibinfo {author} {\bibfnamefont {A.}~\bibnamefont {Ekstr\"om}}, \bibinfo {author} {\bibfnamefont {C.}~\bibnamefont {Forss\'en}}, \bibinfo {author} {\bibfnamefont {R.~J.}\ \bibnamefont {Furnstahl}}, \bibinfo {author} {\bibfnamefont {J.~A.}\ \bibnamefont {Melendez}},\ and\ \bibinfo {author} {\bibfnamefont {D.~R.}\ \bibnamefont {Phillips}},\ }\bibfield  {title} {\bibinfo {title} {Rigorous constraints on three-nucleon forces in chiral effective field theory from fast and accurate calculations of few-body observables},\ }\href {https://doi.org/10.1103/PhysRevC.104.064001} {\bibfield  {journal} {\bibinfo  {journal} {Phys. Rev. C}\ }\textbf {\bibinfo {volume} {104}},\ \bibinfo {pages} {064001} (\bibinfo {year} {2021})}\BibitemShut {NoStop}%
\bibitem [{\citenamefont {Svensson}\ \emph {et~al.}(2023{\natexlab{a}})\citenamefont {Svensson}, \citenamefont {Ekström},\ and\ \citenamefont {Forssén}}]{Svensson_2023}%
  \BibitemOpen
  \bibfield  {author} {\bibinfo {author} {\bibfnamefont {I.}~\bibnamefont {Svensson}}, \bibinfo {author} {\bibfnamefont {A.}~\bibnamefont {Ekström}},\ and\ \bibinfo {author} {\bibfnamefont {C.}~\bibnamefont {Forssén}},\ }\bibfield  {title} {\bibinfo {title} {Bayesian estimation of the low-energy constants up to fourth order in the nucleon-nucleon sector of chiral effective field theory},\ }\bibfield  {journal} {\bibinfo  {journal} {Physical Review C}\ }\textbf {\bibinfo {volume} {107}},\ \href {https://doi.org/10.1103/physrevc.107.014001} {10.1103/physrevc.107.014001} (\bibinfo {year} {2023}{\natexlab{a}})\BibitemShut {NoStop}%
\bibitem [{\citenamefont {Svensson}\ \emph {et~al.}(2023{\natexlab{b}})\citenamefont {Svensson}, \citenamefont {Ekström},\ and\ \citenamefont {Forssén}}]{svensson2023inference}%
  \BibitemOpen
  \bibfield  {author} {\bibinfo {author} {\bibfnamefont {I.}~\bibnamefont {Svensson}}, \bibinfo {author} {\bibfnamefont {A.}~\bibnamefont {Ekström}},\ and\ \bibinfo {author} {\bibfnamefont {C.}~\bibnamefont {Forssén}},\ }\href@noop {} {\bibinfo {title} {Inference of the low-energy constants in delta-full chiral effective field theory including a correlated truncation error}} (\bibinfo {year} {2023}{\natexlab{b}}),\ \Eprint {https://arxiv.org/abs/2304.02004} {arXiv:2304.02004 [nucl-th]} \BibitemShut {NoStop}%
\bibitem [{\citenamefont {Machleidt}\ and\ \citenamefont {Entem}(2011)}]{Machleidt_2011}%
  \BibitemOpen
  \bibfield  {author} {\bibinfo {author} {\bibfnamefont {R.}~\bibnamefont {Machleidt}}\ and\ \bibinfo {author} {\bibfnamefont {D.~R.}\ \bibnamefont {Entem}},\ }\bibfield  {title} {\bibinfo {title} {Chiral effective field theory and nuclear forces},\ }\href {https://doi.org/10.1016/j.physrep.2011.02.001} {\bibfield  {journal} {\bibinfo  {journal} {Physics Reports}\ }\textbf {\bibinfo {volume} {503}},\ \bibinfo {pages} {1} (\bibinfo {year} {2011})}\BibitemShut {NoStop}%
\bibitem [{\citenamefont {Dytrych}\ \emph {et~al.}(2007)\citenamefont {Dytrych}, \citenamefont {Sviratcheva}, \citenamefont {Bahri}, \citenamefont {Draayer},\ and\ \citenamefont {Vary}}]{DytrychSBDV_PRL07}%
  \BibitemOpen
  \bibfield  {author} {\bibinfo {author} {\bibfnamefont {T.}~\bibnamefont {Dytrych}}, \bibinfo {author} {\bibfnamefont {K.~D.}\ \bibnamefont {Sviratcheva}}, \bibinfo {author} {\bibfnamefont {C.}~\bibnamefont {Bahri}}, \bibinfo {author} {\bibfnamefont {J.~P.}\ \bibnamefont {Draayer}},\ and\ \bibinfo {author} {\bibfnamefont {J.~P.}\ \bibnamefont {Vary}},\ }\bibfield  {title} {\bibinfo {title} {{Evidence for Symplectic Symmetry in ab initio No-core-shell-model Results for Light Nuclei}},\ }\href@noop {} {\bibfield  {journal} {\bibinfo  {journal} {Phys. Rev. Lett.}\ }\textbf {\bibinfo {volume} {98}},\ \bibinfo {pages} {162503} (\bibinfo {year} {2007})}\BibitemShut {NoStop}%
\bibitem [{\citenamefont {Launey}\ \emph {et~al.}(2016)\citenamefont {Launey}, \citenamefont {Dytrych},\ and\ \citenamefont {Draayer}}]{LauneyDD16}%
  \BibitemOpen
  \bibfield  {author} {\bibinfo {author} {\bibfnamefont {K.~D.}\ \bibnamefont {Launey}}, \bibinfo {author} {\bibfnamefont {T.}~\bibnamefont {Dytrych}},\ and\ \bibinfo {author} {\bibfnamefont {J.~P.}\ \bibnamefont {Draayer}},\ }\bibfield  {title} {\bibinfo {title} {{Symmetry-guided large-scale shell-model theory}},\ }\href {https://doi.org/10.1016/j.ppnp.2016.02.001} {\bibfield  {journal} {\bibinfo  {journal} {Prog. Part. Nucl. Phys.}\ }\textbf {\bibinfo {volume} {89}},\ \bibinfo {pages} {101 (review)} (\bibinfo {year} {2016})}\BibitemShut {NoStop}%
\bibitem [{\citenamefont {Dytrych}\ \emph {et~al.}(2020)\citenamefont {Dytrych}, \citenamefont {Launey}, \citenamefont {Draayer}, \citenamefont {Rowe}, \citenamefont {Wood}, \citenamefont {Rosensteel}, \citenamefont {Bahri}, \citenamefont {Langr},\ and\ \citenamefont {Baker}}]{DytrychLDRWRBB20}%
  \BibitemOpen
  \bibfield  {author} {\bibinfo {author} {\bibfnamefont {T.}~\bibnamefont {Dytrych}}, \bibinfo {author} {\bibfnamefont {K.~D.}\ \bibnamefont {Launey}}, \bibinfo {author} {\bibfnamefont {J.~P.}\ \bibnamefont {Draayer}}, \bibinfo {author} {\bibfnamefont {D.~J.}\ \bibnamefont {Rowe}}, \bibinfo {author} {\bibfnamefont {J.~L.}\ \bibnamefont {Wood}}, \bibinfo {author} {\bibfnamefont {G.}~\bibnamefont {Rosensteel}}, \bibinfo {author} {\bibfnamefont {C.}~\bibnamefont {Bahri}}, \bibinfo {author} {\bibfnamefont {D.}~\bibnamefont {Langr}},\ and\ \bibinfo {author} {\bibfnamefont {R.~B.}\ \bibnamefont {Baker}},\ }\bibfield  {title} {\bibinfo {title} {Physics of nuclei: Key role of an emergent symmetry},\ }\href {https://doi.org/10.1103/PhysRevLett.124.042501} {\bibfield  {journal} {\bibinfo  {journal} {Phys. Rev. Lett.}\ }\textbf {\bibinfo {volume} {124}},\ \bibinfo {pages} {042501} (\bibinfo {year} {2020})}\BibitemShut {NoStop}%
\bibitem [{\citenamefont {Ekstr{\"o}m}\ \emph {et~al.}(2013)\citenamefont {Ekstr{\"o}m}, \citenamefont {Baardsen}, \citenamefont {Forss{\'e}n}, \citenamefont {Hagen}, \citenamefont {Hjorth-Jensen}, \citenamefont {Jansen}, \citenamefont {Machleidt}, \citenamefont {Nazarewicz} \emph {et~al.}}]{Ekstrom13}%
  \BibitemOpen
  \bibfield  {author} {\bibinfo {author} {\bibfnamefont {A.}~\bibnamefont {Ekstr{\"o}m}}, \bibinfo {author} {\bibfnamefont {G.}~\bibnamefont {Baardsen}}, \bibinfo {author} {\bibfnamefont {C.}~\bibnamefont {Forss{\'e}n}}, \bibinfo {author} {\bibfnamefont {G.}~\bibnamefont {Hagen}}, \bibinfo {author} {\bibfnamefont {M.}~\bibnamefont {Hjorth-Jensen}}, \bibinfo {author} {\bibfnamefont {G.~R.}\ \bibnamefont {Jansen}}, \bibinfo {author} {\bibfnamefont {R.}~\bibnamefont {Machleidt}}, \bibinfo {author} {\bibfnamefont {W.}~\bibnamefont {Nazarewicz}}, \emph {et~al.},\ }\bibfield  {title} {\bibinfo {title} {{An optimized chiral nucleon-nucleon interaction at next-to-next-to-leading order}},\ }\href@noop {} {\bibfield  {journal} {\bibinfo  {journal} {Phys. Rev. Lett.}\ }\textbf {\bibinfo {volume} {110}},\ \bibinfo {pages} {192502} (\bibinfo {year} {2013})}\BibitemShut {NoStop}%
\bibitem [{\citenamefont {Launey}\ \emph {et~al.}(2021)\citenamefont {Launey}, \citenamefont {Mercenne},\ and\ \citenamefont {Dytrych}}]{LauneyMD_ARNPS21}%
  \BibitemOpen
  \bibfield  {author} {\bibinfo {author} {\bibfnamefont {K.~D.}\ \bibnamefont {Launey}}, \bibinfo {author} {\bibfnamefont {A.}~\bibnamefont {Mercenne}},\ and\ \bibinfo {author} {\bibfnamefont {T.}~\bibnamefont {Dytrych}},\ }\bibfield  {title} {\bibinfo {title} {Nuclear dynamics and reactions in the ab initio symmetry-adapted framework},\ }\href {https://doi.org/10.1146/annurev-nucl-102419-033316} {\bibfield  {journal} {\bibinfo  {journal} {Annu. Rev. Nucl. Part. Sci.}\ }\textbf {\bibinfo {volume} {71}},\ \bibinfo {pages} {253} (\bibinfo {year} {2021})}\BibitemShut {NoStop}%
\bibitem [{\citenamefont {Navr\'{a}til}\ \emph {et~al.}(2000)\citenamefont {Navr\'{a}til}, \citenamefont {Vary},\ and\ \citenamefont {Barrett}}]{NavratilVB00}%
  \BibitemOpen
  \bibfield  {author} {\bibinfo {author} {\bibfnamefont {P.}~\bibnamefont {Navr\'{a}til}}, \bibinfo {author} {\bibfnamefont {J.~P.}\ \bibnamefont {Vary}},\ and\ \bibinfo {author} {\bibfnamefont {B.~R.}\ \bibnamefont {Barrett}},\ }\bibfield  {title} {\bibinfo {title} {{Properties of $^{12}$C in the {\it Ab Initio} Nuclear Shell Model}},\ }\href@noop {} {\bibfield  {journal} {\bibinfo  {journal} {Phys. Rev. Lett.}\ }\textbf {\bibinfo {volume} {84}},\ \bibinfo {pages} {5728} (\bibinfo {year} {2000})}\BibitemShut {NoStop}%
\bibitem [{\citenamefont {Barrett}\ \emph {et~al.}(2013)\citenamefont {Barrett}, \citenamefont {Navr\'{a}til},\ and\ \citenamefont {Vary}}]{BarrettNV13}%
  \BibitemOpen
  \bibfield  {author} {\bibinfo {author} {\bibfnamefont {B.~R.}\ \bibnamefont {Barrett}}, \bibinfo {author} {\bibfnamefont {P.}~\bibnamefont {Navr\'{a}til}},\ and\ \bibinfo {author} {\bibfnamefont {J.~P.}\ \bibnamefont {Vary}},\ }\bibfield  {title} {\bibinfo {title} {Ab initio no core shell model},\ }\href@noop {} {\bibfield  {journal} {\bibinfo  {journal} {Prog. Part. Nucl. Phys.}\ }\textbf {\bibinfo {volume} {69}},\ \bibinfo {pages} {131} (\bibinfo {year} {2013})}\BibitemShut {NoStop}%
\bibitem [{\citenamefont {Launey}\ \emph {et~al.}(2020)\citenamefont {Launey}, \citenamefont {Dytrych}, \citenamefont {Sargsyan}, \citenamefont {Baker},\ and\ \citenamefont {Draayer}}]{LauneyDSBD20}%
  \BibitemOpen
  \bibfield  {author} {\bibinfo {author} {\bibfnamefont {K.~D.}\ \bibnamefont {Launey}}, \bibinfo {author} {\bibfnamefont {T.}~\bibnamefont {Dytrych}}, \bibinfo {author} {\bibfnamefont {G.~H.}\ \bibnamefont {Sargsyan}}, \bibinfo {author} {\bibfnamefont {R.~B.}\ \bibnamefont {Baker}},\ and\ \bibinfo {author} {\bibfnamefont {J.~P.}\ \bibnamefont {Draayer}},\ }\bibfield  {title} {\bibinfo {title} {{Emergent symplectic symmetry in atomic nuclei: \textit{Ab initio} symmetry-adapted no-core shell model}},\ }\href {https://doi.org/10.1140/epjst/e2020-000178-3} {\bibfield  {journal} {\bibinfo  {journal} {Eur. Phys. J. Spec. Top.}\ }\textbf {\bibinfo {volume} {229}},\ \bibinfo {pages} {2429} (\bibinfo {year} {2020})}\BibitemShut {NoStop}%
\bibitem [{\citenamefont {Bahri}\ and\ \citenamefont {Rowe}(2000)}]{BahriR00}%
  \BibitemOpen
  \bibfield  {author} {\bibinfo {author} {\bibfnamefont {C.}~\bibnamefont {Bahri}}\ and\ \bibinfo {author} {\bibfnamefont {D.~J.}\ \bibnamefont {Rowe}},\ }\bibfield  {title} {\bibinfo {title} {Su(3)quasi-dynamical symmetry as an organizational mechanism for generating nuclear rotational motions},\ }\href@noop {} {\bibfield  {journal} {\bibinfo  {journal} {Nucl. Phys. A}\ }\textbf {\bibinfo {volume} {662}},\ \bibinfo {pages} {125} (\bibinfo {year} {2000})}\BibitemShut {NoStop}%
\bibitem [{\citenamefont {Dytrych}\ \emph {et~al.}(2013)\citenamefont {Dytrych}, \citenamefont {Launey}, \citenamefont {Draayer}, \citenamefont {Maris}, \citenamefont {Vary}, \citenamefont {Saule}, \citenamefont {Catalyurek}, \citenamefont {Sosonkina}, \citenamefont {Langr},\ and\ \citenamefont {Caprio}}]{dytrychlmcdvl_prl12}%
  \BibitemOpen
  \bibfield  {author} {\bibinfo {author} {\bibfnamefont {T.}~\bibnamefont {Dytrych}}, \bibinfo {author} {\bibfnamefont {K.~D.}\ \bibnamefont {Launey}}, \bibinfo {author} {\bibfnamefont {J.~P.}\ \bibnamefont {Draayer}}, \bibinfo {author} {\bibfnamefont {P.}~\bibnamefont {Maris}}, \bibinfo {author} {\bibfnamefont {J.~P.}\ \bibnamefont {Vary}}, \bibinfo {author} {\bibfnamefont {E.}~\bibnamefont {Saule}}, \bibinfo {author} {\bibfnamefont {U.}~\bibnamefont {Catalyurek}}, \bibinfo {author} {\bibfnamefont {M.}~\bibnamefont {Sosonkina}}, \bibinfo {author} {\bibfnamefont {D.}~\bibnamefont {Langr}},\ and\ \bibinfo {author} {\bibfnamefont {M.~A.}\ \bibnamefont {Caprio}},\ }\bibfield  {title} {\bibinfo {title} {{Collective Modes in Light Nuclei from First Principles}},\ }\href {https://doi.org/10.1103/PhysRevLett.111.252501} {\bibfield  {journal} {\bibinfo  {journal} {Phys. Rev. Lett.}\ }\textbf {\bibinfo {volume} {111}},\ \bibinfo {pages} {252501} (\bibinfo {year} {2013})}\BibitemShut {NoStop}%
\bibitem [{\citenamefont {Dreyfuss}\ \emph {et~al.}(2017)\citenamefont {Dreyfuss}, \citenamefont {Launey}, \citenamefont {Dytrych}, \citenamefont {Draayer}, \citenamefont {Baker}, \citenamefont {Deibel},\ and\ \citenamefont {Bahri}}]{DreyfussLTDBDB16}%
  \BibitemOpen
  \bibfield  {author} {\bibinfo {author} {\bibfnamefont {A.~C.}\ \bibnamefont {Dreyfuss}}, \bibinfo {author} {\bibfnamefont {K.~D.}\ \bibnamefont {Launey}}, \bibinfo {author} {\bibfnamefont {T.}~\bibnamefont {Dytrych}}, \bibinfo {author} {\bibfnamefont {J.~P.}\ \bibnamefont {Draayer}}, \bibinfo {author} {\bibfnamefont {R.~B.}\ \bibnamefont {Baker}}, \bibinfo {author} {\bibfnamefont {C.~M.}\ \bibnamefont {Deibel}},\ and\ \bibinfo {author} {\bibfnamefont {C.}~\bibnamefont {Bahri}},\ }\bibfield  {title} {\bibinfo {title} {{Understanding emergent collectivity and clustering in nuclei from a symmetry-based no-core shell-model perspective}},\ }\href@noop {} {\bibfield  {journal} {\bibinfo  {journal} {Phys. Rev. C}\ }\textbf {\bibinfo {volume} {95}},\ \bibinfo {pages} {044312} (\bibinfo {year} {2017})}\BibitemShut {NoStop}%
\bibitem [{\citenamefont {Tobin}\ \emph {et~al.}(2014)\citenamefont {Tobin}, \citenamefont {Ferriss}, \citenamefont {Launey}, \citenamefont {Dytrych}, \citenamefont {Draayer},\ and\ \citenamefont {Bahri}}]{TobinFLDDB14}%
  \BibitemOpen
  \bibfield  {author} {\bibinfo {author} {\bibfnamefont {G.~K.}\ \bibnamefont {Tobin}}, \bibinfo {author} {\bibfnamefont {M.~C.}\ \bibnamefont {Ferriss}}, \bibinfo {author} {\bibfnamefont {K.~D.}\ \bibnamefont {Launey}}, \bibinfo {author} {\bibfnamefont {T.}~\bibnamefont {Dytrych}}, \bibinfo {author} {\bibfnamefont {J.~P.}\ \bibnamefont {Draayer}},\ and\ \bibinfo {author} {\bibfnamefont {C.}~\bibnamefont {Bahri}},\ }\bibfield  {title} {\bibinfo {title} {{Symplectic No-core Shell-model Approach to Intermediate-mass Nuclei}},\ }\href@noop {} {\bibfield  {journal} {\bibinfo  {journal} {Phys. Rev. C}\ }\textbf {\bibinfo {volume} {89}},\ \bibinfo {pages} {034312} (\bibinfo {year} {2014})}\BibitemShut {NoStop}%
\bibitem [{\citenamefont {Gasparyan}\ and\ \citenamefont {Epelbaum}(2023)}]{Gasparyan_2023}%
  \BibitemOpen
  \bibfield  {author} {\bibinfo {author} {\bibfnamefont {A.~M.}\ \bibnamefont {Gasparyan}}\ and\ \bibinfo {author} {\bibfnamefont {E.}~\bibnamefont {Epelbaum}},\ }\bibfield  {title} {\bibinfo {title} {Renormalization of nuclear chiral effective field theory with nonperturbative leading-order interactions},\ }\bibfield  {journal} {\bibinfo  {journal} {Physical Review C}\ }\textbf {\bibinfo {volume} {107}},\ \href {https://doi.org/10.1103/physrevc.107.044002} {10.1103/physrevc.107.044002} (\bibinfo {year} {2023})\BibitemShut {NoStop}%
\bibitem [{\citenamefont {Chang}\ \emph {et~al.}(2018)\citenamefont {Chang}, \citenamefont {Nicholson}, \citenamefont {Rinaldi}, \citenamefont {Berkowitz}, \citenamefont {Garron}, \citenamefont {Brantley}, \citenamefont {Monge-Camacho}, \citenamefont {Monahan}, \citenamefont {Bouchard}, \citenamefont {Clark}, \citenamefont {Joó}, \citenamefont {Kurth}, \citenamefont {Orginos}, \citenamefont {Vranas},\ and\ \citenamefont {Walker-Loud}}]{Chang_2018}%
  \BibitemOpen
  \bibfield  {author} {\bibinfo {author} {\bibfnamefont {C.~C.}\ \bibnamefont {Chang}}, \bibinfo {author} {\bibfnamefont {A.~N.}\ \bibnamefont {Nicholson}}, \bibinfo {author} {\bibfnamefont {E.}~\bibnamefont {Rinaldi}}, \bibinfo {author} {\bibfnamefont {E.}~\bibnamefont {Berkowitz}}, \bibinfo {author} {\bibfnamefont {N.}~\bibnamefont {Garron}}, \bibinfo {author} {\bibfnamefont {D.~A.}\ \bibnamefont {Brantley}}, \bibinfo {author} {\bibfnamefont {H.}~\bibnamefont {Monge-Camacho}}, \bibinfo {author} {\bibfnamefont {C.~J.}\ \bibnamefont {Monahan}}, \bibinfo {author} {\bibfnamefont {C.}~\bibnamefont {Bouchard}}, \bibinfo {author} {\bibfnamefont {M.~A.}\ \bibnamefont {Clark}}, \bibinfo {author} {\bibfnamefont {B.}~\bibnamefont {Joó}}, \bibinfo {author} {\bibfnamefont {T.}~\bibnamefont {Kurth}}, \bibinfo {author} {\bibfnamefont {K.}~\bibnamefont {Orginos}}, \bibinfo {author} {\bibfnamefont {P.}~\bibnamefont {Vranas}},\ and\ \bibinfo {author} {\bibfnamefont {A.}~\bibnamefont {Walker-Loud}},\ }\bibfield  {title}
  {\bibinfo {title} {A per-cent-level determination of the nucleon axial coupling from quantum chromodynamics},\ }\href {https://doi.org/10.1038/s41586-018-0161-8} {\bibfield  {journal} {\bibinfo  {journal} {Nature}\ }\textbf {\bibinfo {volume} {558}},\ \bibinfo {pages} {91–94} (\bibinfo {year} {2018})}\BibitemShut {NoStop}%
\bibitem [{\citenamefont {Dreyfuss}\ \emph {et~al.}(2013)\citenamefont {Dreyfuss}, \citenamefont {Launey}, \citenamefont {Dytrych}, \citenamefont {Draayer},\ and\ \citenamefont {Bahri}}]{DreyfussLTDB13}%
  \BibitemOpen
  \bibfield  {author} {\bibinfo {author} {\bibfnamefont {A.~C.}\ \bibnamefont {Dreyfuss}}, \bibinfo {author} {\bibfnamefont {K.~D.}\ \bibnamefont {Launey}}, \bibinfo {author} {\bibfnamefont {T.}~\bibnamefont {Dytrych}}, \bibinfo {author} {\bibfnamefont {J.~P.}\ \bibnamefont {Draayer}},\ and\ \bibinfo {author} {\bibfnamefont {C.}~\bibnamefont {Bahri}},\ }\bibfield  {title} {\bibinfo {title} {{Hoyle state and rotational features in Carbon-12 within a no-core shell model framework}},\ }\href@noop {} {\bibfield  {journal} {\bibinfo  {journal} {Phys. Lett. B}\ }\textbf {\bibinfo {volume} {727}},\ \bibinfo {pages} {511} (\bibinfo {year} {2013})}\BibitemShut {NoStop}%
\bibitem [{\citenamefont {Baker}\ \emph {et~al.}(2020)\citenamefont {Baker}, \citenamefont {Launey}, \citenamefont {Bacca}, \citenamefont {Dinur},\ and\ \citenamefont {Dytrych}}]{BakerLBND20}%
  \BibitemOpen
  \bibfield  {author} {\bibinfo {author} {\bibfnamefont {R.~B.}\ \bibnamefont {Baker}}, \bibinfo {author} {\bibfnamefont {K.~D.}\ \bibnamefont {Launey}}, \bibinfo {author} {\bibfnamefont {S.}~\bibnamefont {Bacca}}, \bibinfo {author} {\bibfnamefont {N.~N.}\ \bibnamefont {Dinur}},\ and\ \bibinfo {author} {\bibfnamefont {T.}~\bibnamefont {Dytrych}},\ }\bibfield  {title} {\bibinfo {title} {Benchmark calculations of electromagnetic sum rules with a symmetry-adapted basis and hyperspherical harmonics},\ }\href {https://doi.org/10.1103/PhysRevC.102.014320} {\bibfield  {journal} {\bibinfo  {journal} {Phys. Rev. C}\ }\textbf {\bibinfo {volume} {102}},\ \bibinfo {pages} {014320} (\bibinfo {year} {2020})}\BibitemShut {NoStop}%
\bibitem [{\citenamefont {Burrows}\ \emph {et~al.}(2019)\citenamefont {Burrows}, \citenamefont {Elster}, \citenamefont {Weppner}, \citenamefont {Launey}, \citenamefont {Maris}, \citenamefont {Nogga},\ and\ \citenamefont {Popa}}]{BurrowsEWLMNP19}%
  \BibitemOpen
  \bibfield  {author} {\bibinfo {author} {\bibfnamefont {M.}~\bibnamefont {Burrows}}, \bibinfo {author} {\bibfnamefont {C.}~\bibnamefont {Elster}}, \bibinfo {author} {\bibfnamefont {S.~P.}\ \bibnamefont {Weppner}}, \bibinfo {author} {\bibfnamefont {K.~D.}\ \bibnamefont {Launey}}, \bibinfo {author} {\bibfnamefont {P.}~\bibnamefont {Maris}}, \bibinfo {author} {\bibfnamefont {A.}~\bibnamefont {Nogga}},\ and\ \bibinfo {author} {\bibfnamefont {G.}~\bibnamefont {Popa}},\ }\bibfield  {title} {\bibinfo {title} {Ab initio folding potentials for nucleon-nucleus scattering based on no-core shell-model one-body densities},\ }\href {https://doi.org/10.1103/PhysRevC.99.044603} {\bibfield  {journal} {\bibinfo  {journal} {Phys. Rev. C}\ }\textbf {\bibinfo {volume} {99}},\ \bibinfo {pages} {044603} (\bibinfo {year} {2019})}\BibitemShut {NoStop}%
\bibitem [{\citenamefont {Ruotsalainen}\ \emph {et~al.}(2019)\citenamefont {Ruotsalainen}, \citenamefont {Henderson}, \citenamefont {Hackman}, \citenamefont {Sargsyan}, \citenamefont {Launey}, \citenamefont {Saxena}, \citenamefont {Srivastava}, \citenamefont {Stroberg}, \citenamefont {Grahn}, \citenamefont {Pakarinen}, \citenamefont {Ball}, \citenamefont {Julin}, \citenamefont {Greenlees}, \citenamefont {Smallcombe}, \citenamefont {Andreoiu}, \citenamefont {Bernier}, \citenamefont {Bowry}, \citenamefont {Buckner}, \citenamefont {Caballero-Folch}, \citenamefont {Chester}, \citenamefont {Cruz}, \citenamefont {Evitts}, \citenamefont {Frederick}, \citenamefont {Garnsworthy}, \citenamefont {Holl}, \citenamefont {Kurkjian}, \citenamefont {Kisliuk}, \citenamefont {Leach}, \citenamefont {McGee}, \citenamefont {Measures}, \citenamefont {M\"ucher}, \citenamefont {Park}, \citenamefont {Sarazin}, \citenamefont {Smith}, \citenamefont {Southall}, \citenamefont {Starosta}, \citenamefont {Svensson}, \citenamefont {Whitmore},
  \citenamefont {Williams},\ and\ \citenamefont {Wu}}]{Ruotsalainen19}%
  \BibitemOpen
  \bibfield  {author} {\bibinfo {author} {\bibfnamefont {P.}~\bibnamefont {Ruotsalainen}}, \bibinfo {author} {\bibfnamefont {J.}~\bibnamefont {Henderson}}, \bibinfo {author} {\bibfnamefont {G.}~\bibnamefont {Hackman}}, \bibinfo {author} {\bibfnamefont {G.~H.}\ \bibnamefont {Sargsyan}}, \bibinfo {author} {\bibfnamefont {K.~D.}\ \bibnamefont {Launey}}, \bibinfo {author} {\bibfnamefont {A.}~\bibnamefont {Saxena}}, \bibinfo {author} {\bibfnamefont {P.~C.}\ \bibnamefont {Srivastava}}, \bibinfo {author} {\bibfnamefont {S.~R.}\ \bibnamefont {Stroberg}}, \bibinfo {author} {\bibfnamefont {T.}~\bibnamefont {Grahn}}, \bibinfo {author} {\bibfnamefont {J.}~\bibnamefont {Pakarinen}}, \bibinfo {author} {\bibfnamefont {G.~C.}\ \bibnamefont {Ball}}, \bibinfo {author} {\bibfnamefont {R.}~\bibnamefont {Julin}}, \bibinfo {author} {\bibfnamefont {P.~T.}\ \bibnamefont {Greenlees}}, \bibinfo {author} {\bibfnamefont {J.}~\bibnamefont {Smallcombe}}, \bibinfo {author} {\bibfnamefont {C.}~\bibnamefont {Andreoiu}}, \bibinfo {author}
  {\bibfnamefont {N.}~\bibnamefont {Bernier}}, \bibinfo {author} {\bibfnamefont {M.}~\bibnamefont {Bowry}}, \bibinfo {author} {\bibfnamefont {M.}~\bibnamefont {Buckner}}, \bibinfo {author} {\bibfnamefont {R.}~\bibnamefont {Caballero-Folch}}, \bibinfo {author} {\bibfnamefont {A.}~\bibnamefont {Chester}}, \bibinfo {author} {\bibfnamefont {S.}~\bibnamefont {Cruz}}, \bibinfo {author} {\bibfnamefont {L.~J.}\ \bibnamefont {Evitts}}, \bibinfo {author} {\bibfnamefont {R.}~\bibnamefont {Frederick}}, \bibinfo {author} {\bibfnamefont {A.~B.}\ \bibnamefont {Garnsworthy}}, \bibinfo {author} {\bibfnamefont {M.}~\bibnamefont {Holl}}, \bibinfo {author} {\bibfnamefont {A.}~\bibnamefont {Kurkjian}}, \bibinfo {author} {\bibfnamefont {D.}~\bibnamefont {Kisliuk}}, \bibinfo {author} {\bibfnamefont {K.~G.}\ \bibnamefont {Leach}}, \bibinfo {author} {\bibfnamefont {E.}~\bibnamefont {McGee}}, \bibinfo {author} {\bibfnamefont {J.}~\bibnamefont {Measures}}, \bibinfo {author} {\bibfnamefont {D.}~\bibnamefont {M\"ucher}}, \bibinfo
  {author} {\bibfnamefont {J.}~\bibnamefont {Park}}, \bibinfo {author} {\bibfnamefont {F.}~\bibnamefont {Sarazin}}, \bibinfo {author} {\bibfnamefont {J.~K.}\ \bibnamefont {Smith}}, \bibinfo {author} {\bibfnamefont {D.}~\bibnamefont {Southall}}, \bibinfo {author} {\bibfnamefont {K.}~\bibnamefont {Starosta}}, \bibinfo {author} {\bibfnamefont {C.~E.}\ \bibnamefont {Svensson}}, \bibinfo {author} {\bibfnamefont {K.}~\bibnamefont {Whitmore}}, \bibinfo {author} {\bibfnamefont {M.}~\bibnamefont {Williams}},\ and\ \bibinfo {author} {\bibfnamefont {C.~Y.}\ \bibnamefont {Wu}},\ }\bibfield  {title} {\bibinfo {title} {Isospin symmetry in $b(e2)$ values: Coulomb excitation study of $^{21}\mathrm{Mg}$},\ }\href {https://doi.org/10.1103/PhysRevC.99.051301} {\bibfield  {journal} {\bibinfo  {journal} {Phys. Rev. C}\ }\textbf {\bibinfo {volume} {99}},\ \bibinfo {pages} {051301} (\bibinfo {year} {2019})}\BibitemShut {NoStop}%
\bibitem [{\citenamefont {Miller}\ \emph {et~al.}(2022)\citenamefont {Miller}, \citenamefont {Ekstr{\"o}m},\ and\ \citenamefont {Hebeler}}]{Miller22}%
  \BibitemOpen
  \bibfield  {author} {\bibinfo {author} {\bibfnamefont {S.}~\bibnamefont {Miller}}, \bibinfo {author} {\bibfnamefont {A.}~\bibnamefont {Ekstr{\"o}m}},\ and\ \bibinfo {author} {\bibfnamefont {K.}~\bibnamefont {Hebeler}},\ }\bibfield  {title} {\bibinfo {title} {Neutron-deuteron scattering cross-sections with chiral $ nn $ interactions using wave-packet continuum discretization},\ }\href@noop {} {\bibfield  {journal} {\bibinfo  {journal} {arXiv preprint arXiv:2201.09600}\ } (\bibinfo {year} {2022})},\ \bibinfo {note} {arXiv:2201.09600}\BibitemShut {NoStop}%
\bibitem [{\citenamefont {Becker}\ \emph {et~al.}(2024{\natexlab{a}})\citenamefont {Becker}, \citenamefont {Launey}, \citenamefont {Ekström}, \citenamefont {Sargsyan}, \citenamefont {Mumma}, \citenamefont {Dytrych}, \citenamefont {Langr},\ and\ \citenamefont {Draayer}}]{Becker_CGS_2023}%
  \BibitemOpen
  \bibfield  {author} {\bibinfo {author} {\bibfnamefont {K.~S.}\ \bibnamefont {Becker}}, \bibinfo {author} {\bibfnamefont {K.~D.}\ \bibnamefont {Launey}}, \bibinfo {author} {\bibfnamefont {A.}~\bibnamefont {Ekström}}, \bibinfo {author} {\bibfnamefont {G.~H.}\ \bibnamefont {Sargsyan}}, \bibinfo {author} {\bibfnamefont {D.}~\bibnamefont {Mumma}}, \bibinfo {author} {\bibfnamefont {T.}~\bibnamefont {Dytrych}}, \bibinfo {author} {\bibfnamefont {D.}~\bibnamefont {Langr}},\ and\ \bibinfo {author} {\bibfnamefont {J.~P.}\ \bibnamefont {Draayer}},\ }\bibfield  {title} {\bibinfo {title} {Understanding the effect of chiral nn parametrization on nuclear shapes from an ab initio perspective},\ }\href@noop {} {\bibfield  {journal} {\bibinfo  {journal} {Eur. Phys. J. (in review)}\ } (\bibinfo {year} {2024}{\natexlab{a}})}\BibitemShut {NoStop}%
\bibitem [{\citenamefont {Becker}\ \emph {et~al.}(2024{\natexlab{b}})\citenamefont {Becker}, \citenamefont {Launey}, \citenamefont {Ekström}, \citenamefont {Dytrych}, \citenamefont {Langr}, \citenamefont {Sargsyan},\ and\ \citenamefont {Draayer}}]{Becker_PRL_2024}%
  \BibitemOpen
  \bibfield  {author} {\bibinfo {author} {\bibfnamefont {K.~S.}\ \bibnamefont {Becker}}, \bibinfo {author} {\bibfnamefont {K.~D.}\ \bibnamefont {Launey}}, \bibinfo {author} {\bibfnamefont {A.}~\bibnamefont {Ekström}}, \bibinfo {author} {\bibfnamefont {T.}~\bibnamefont {Dytrych}}, \bibinfo {author} {\bibfnamefont {D.}~\bibnamefont {Langr}}, \bibinfo {author} {\bibfnamefont {G.~H.}\ \bibnamefont {Sargsyan}},\ and\ \bibinfo {author} {\bibfnamefont {J.~P.}\ \bibnamefont {Draayer}},\ }\bibfield  {title} {\bibinfo {title} {The impact of high-energy physics on long-ranged nuclear collectivity},\ }\href@noop {} {\bibfield  {journal} {\bibinfo  {journal} {To be submitted}\ } (\bibinfo {year} {2024}{\natexlab{b}})}\BibitemShut {NoStop}%
\bibitem [{\citenamefont {Sargsyan}\ \emph {et~al.}(2021)\citenamefont {Sargsyan}, \citenamefont {Launey} \emph {et~al.}}]{Sargsyan_A8}%
  \BibitemOpen
  \bibfield  {author} {\bibinfo {author} {\bibfnamefont {G.~H.}\ \bibnamefont {Sargsyan}}, \bibinfo {author} {\bibfnamefont {K.~D.}\ \bibnamefont {Launey}}, \emph {et~al.},\ }\bibfield  {title} {\bibinfo {title} {{Impact of clustering on the 8Li beta decay and recoil form factors}},\ }\href@noop {} {\bibfield  {journal} {\bibinfo  {journal} {Phys. Rev. Lett. (submitted); arXiv:2107.10389}\ } (\bibinfo {year} {2021})}\BibitemShut {NoStop}%
\bibitem [{\citenamefont {Becker}\ \emph {et~al.}(2023)\citenamefont {Becker}, \citenamefont {Launey}, \citenamefont {Ekstrom},\ and\ \citenamefont {Dytrych}}]{BeckerLED22}%
  \BibitemOpen
  \bibfield  {author} {\bibinfo {author} {\bibfnamefont {K.~S.}\ \bibnamefont {Becker}}, \bibinfo {author} {\bibfnamefont {K.~D.}\ \bibnamefont {Launey}}, \bibinfo {author} {\bibfnamefont {A.}~\bibnamefont {Ekstrom}},\ and\ \bibinfo {author} {\bibfnamefont {T.}~\bibnamefont {Dytrych}},\ }\bibfield  {title} {\bibinfo {title} {Ab initio symmetry-adapted emulator for studying emergent collectivity and clustering in nuclei},\ }\href {https://doi.org/10.3389/fphy.2023.1064601} {\bibfield  {journal} {\bibinfo  {journal} {Front. Phys.}\ }\textbf {\bibinfo {volume} {11}},\ \bibinfo {pages} {1064601} (\bibinfo {year} {2023})}\BibitemShut {NoStop}%
\end{thebibliography}%

\end{document}